\documentclass[extra,mreferee]{gji}

\usepackage{timet}
\usepackage[latin1]{inputenc}
\usepackage{amsmath, amstext, amsfonts}
\usepackage{gensymb}
\usepackage{graphicx}
\usepackage{array, color, colortbl}
\usepackage{gji_extra}
\usepackage[section]{placeins}

\makeatletter
\let\@tabclassz =\@classz
\makeatother 

\usepackage[colorlinks,linkcolor=blue,urlcolor=blue,citecolor=blue]{hyperref}
\usepackage{url}

\usepackage{lineno}
\title{Changes in core-mantle boundary heat flux patterns throughout the supercontinent cycle}
\author[J. Dannberg et al.]
  {Juliane Dannberg$^1$ \and Rene Gassmoeller$^1$ \and Daniele Thallner$^1$ \and Frederick LaCombe$^1$ \and Courtney Sprain$^1$ \\
\small $^1$Department of Geological Sciences, University of Florida, Gainesville, USA
  }
\date{}
\pagerange{\pageref{firstpage}--\pageref{lastpage}}
\volume{}
\pubyear{}

\begin{document}

\label{firstpage}

\maketitle

\begin{abstract}
The Earth's magnetic field is generated by a dynamo in the outer core and is crucial for shielding our planet from harmful radiation. Despite the established importance of the core-mantle boundary heat flux as driver for the dynamo, open questions remain about how heat flux heterogeneities affect the magnetic field. Here, we explore the distribution of core-mantle boundary heat flux on Earth and its changes over time using compressible global 3-D mantle convection models in the geodynamic modeling software ASPECT. We discuss the use of the consistent boundary flux method as a tool to more accurately compute boundary heat fluxes in finite element simulations and the workflow to provide the computed heat flux patterns as boundary conditions in geodynamo simulations.
Our models use a plate reconstruction throughout the last 1 billion years---encompassing the complete supercontinent cycle---to determine the location and sinking speed of subducted plates. The results show how mantle upwellings and downwellings create localized heat flux anomalies at the core-mantle boundary that can vary drastically over Earth's history and depend on the properties and evolution of the lowermost mantle as well as the surface subduction zone configuration. 
The distribution of hot and cold structures at the core-mantle boundary changes throughout the supercontinent cycle in terms of location, shape and number, indicating that these structures fluctuate and might have looked very differently in Earth's past. 
We estimate the resulting amplitude of spatial heat flux variations, expressed by the ratio of peak-to-peak amplitude to average heat flux, $q*$, to be at least 2. However, depending on the material properties and the adiabatic heat flux out of the core, $q*$ can easily reach values $>30$. For a given set of material properties, $q*$ generally varies by 30-50\% over time.
Our results have implications for understanding the Earth's thermal evolution and the stability of its magnetic field over geological timescales. They provide insights into the potential effects of the mantle on the magnetic field and pave the way for further exploring questions about the nucleation of the inner core and the past state of the lowermost mantle.
\end{abstract}


\section{Introduction}

The heat flux out of the Earth's outer core into the overlying mantle is one of the drivers for the geodynamo responsible for generating Earth's magnetic field. 
Past modeling studies have demonstrated that both the magnitude and spatial variability of this heat flux have a strong impact on the convection patterns in the outer core and the resulting magnetic field. 
Heterogeneous mantle forcing can organize flow near the top of the core \citep{mound2019regional}, introducing non-zonal structure into the time-averaged magnetic field and giving it a morphology and secular variation matching Earth's modern field \citep{mound2023longitudinal}. 
Geodynamo models also suggest that the amplitude and pattern of the heat flux heterogeneity across the core-mantle boundary (CMB), particularly near the equator, affect the average timing between polarity reversals of the magnetic field \citep{glatzmaier1999role, olson2010geodynamo, olson2014magnetic}, the deviations of the time-averaged field from a geocentric axial dipole \citep{bloxham2000sensitivity, heimpel2013testing}, the polarity transition paths during reversals and excursions \citep{kutzner2004simulated}, the field strength \citep{takahashi2008effects}, and the location of intense geomagnetic flux patches \citep{olson2002time, gubbins2007correlation, sahoo2020response}.
For example, simulations show that locations of minimum field intensity are clearly correlated with above average heat flux patterns at the CMB \citep{korte2022indicators}. 
Furthermore, core-mantle interactions are a major influence on the secular variation of the magnetic field computed in geodynamo simulations \citep{bloxham2000effect}. 

In addition, paleomagnetic data reveal variations in the magnetic field on much longer time scales ($>1$~Myr) than characteristic for circulation in the outer core (centuries), instead matching typical time scales of mantle convection.
For example, there is evidence that the frequency of magnetic field reversals varies in a periodic manner, with Phanerozoic superchrons ($>10$~Myr periods with few to no magnetic reversals) occurring roughly every 200~Myr \citep{biggin2012possible}, often being preceded by extended periods of hyper-reversal \cite[$\sim$10 reversals/Myr;][]{biggin2012possible, meert2016rapid}.
\citet{olson2015mantle} find an inverse correlation between the reversal frequency and the activity of Large Igneous Provinces for the past 160~Myrs of Earth's history, suggesting a link between plume activity and the geodynamo.
The changes in reversal frequency over time are suggested to correlate with changes in magnetic field strength, with lower field values occurring during periods of high reversal frequency, representing a highly unstable state of the magnetic field, and stronger values occurring during extended periods of non-reversal, representing a more stable state \citep{cox1968lengths}.
This apparent coupling between Earth's mantle and core is also evidenced by the correlation of the magnetic field calculated from geodynamo models using the lowermost mantle seismic velocity patterns as boundary condition with the present-day geomagnetic field \citep{gubbins2007correlation}.

These observations and modeling results suggest that the influence of the mantle is reflected in the paleomagnetic record. 
However, understanding this relationship remains difficult for several reasons. 
Although we have reasonable estimates of the present-day heat flux at the core-mantle boundary and seismic tomography models offer insights into the distribution of hot and cold material along this boundary, heat flux patterns have likely undergone substantial changes throughout Earth's history.
Unfortunately, we lack a direct record of these past variations. 
Therefore, the magnitude of magnetic field variations caused by mantle convection is currently not well constrained, posing challenges for interpreting the paleomagnetic record.
In addition, most dynamo simulations of Earth's outer core that have taken into account a heterogeneous heat flux across the CMB have either used simplified patterns based on low degree and order (usually $\leq$2) spherical harmonics \cite[i.e.,][]{bloxham2000sensitivity, bloxham2002time, olson2002time, kutzner2004simulated, aubert2007detecting, takahashi2008effects, sreenivasan2009dynamo, olson2010geodynamo, heimpel2013testing, olson2014magnetic, hori2014ancient, amit2015lower, sahoo2016dynamos, olson2017dynamo, christensen2018geodynamo, meduri2021numerical} or have used seismic tomography \cite[e.g.,][]{Masters1996shear} to infer the heat flux \cite[i.e.,][]{glatzmaier1999role, bloxham2000effect, olson2002time, christensen2003secular, kutzner2004simulated, aubert2007detecting, gubbins2007correlation, aubert2008thermochemical, amit2009mantle, olson2010geodynamo, gubbins2011melting, sreenivasan2011mantle, olson2014magnetic, mound2017heat, christensen2018geodynamo, terra2019preferred, sahoo2020response, mound2023longitudinal}, \cite[for a review, see][]{amit2015towards}. 
Both types of patterns only represent the large-scale variations in heat flux, but do not capture smaller-scale variations or strong lateral gradients. 
While imposed heat flux heterogeneities proportional to seismic wave-speed anomalies better represent Earth, they still neglect non-thermal sources likely to contribute to the tomographic pattern. Efforts have been made to address this challenge \citep{amit2009mantle, amit2012mantle, amit2015numerical, choblet2023inferring}, and, as an alternative to seismic tomography, geodynamo models have employed CMB heat flux patterns from a the present-day state of a mantle convection model to capture the relevant physical processes \citep{olson2015core}. 
Nevertheless, even for the present-day, some of the complexities of lower mantle dynamics are not captured in the heat flux patterns used in most geodynamo simulations. 

To our knowledge there has only been one study that has incorporated CMB heat flux patterns corresponding to different times in Earth's history: \citet{olson2013controls} impose a lower mantle history based on time-dependent convection going back to 330~Ma as boundary condition to their geodynamo. They find that these models more readily explain the slow variations in reversal frequency in the Phanerozoic Geomagnetic Polarity Time Scale than models with a heterogeneity pattern that does not change.
However, they include only the largest scale components of thermal core-mantle
interaction by truncating the CMB heat flux pattern at spherical harmonic degree 4 and therefore ignore smaller scale thermal perturbations such as plume formation. In addition, they rely on the heat flux of one specific mantle model \citep{ZHANG2011205} and therefore can not take into account the uncertainty of the CMB heat flux related to uncertainties in lowermost mantle material properties and chemical composition.
%Therefore, many questions about the effect of core--mantle interaction on the magnetic field remain unresolved.  

A deeper understanding of the changes in heat flux patterns at the core-mantle boundary over time and their influence on the geodynamo is essential to determine which observed magnetic field changes can be attributed to mantle convection and which require alternative mechanisms, such as the nucleation of the Earth's inner core.
To constrain these variations, we can use models of mantle convection as a link between available surface observations and the CMB heat flux pattern. 
Subducted slabs sinking down towards the base of the mantle and forming cold areas above the CMB cause a large heat flux out of the core. In addition, they push hotter material together in the regions between downwellings, creating hot regions with a lower CMB heat flux. We therefore expect characteristic changes in CMB heat flux patterns throughout the supercontinent cycle.
Reconstructions of the motion of the tectonic plates at the Earth's surface and their subduction history throughout this cycle can be useful tools to constrain these patterns and their temporal variations. 

\subsection{Temporal and spatial variations in core-mantle boundary heat flux}

Global 3D mantle convection models have been used for several decades to provide insights into spatial and temporal variations of the CMB heat flux. 
These simulations offer a view of how the temperature distribution in the Earth's mantle evolves over time, and when Earth's plate motion history is imposed as boundary condition, they can successfully reproduce the locations of hot regions below Africa and the Pacific inferred from seismic tomography for the present-day Earth (the Large Low Shear Velocity Provinces, or LLSVPs)  \citep{mcnamara2005thermochemical, bull2009synthetic, Zhang2010, davies2012reconciling, bower2013lower, bull2014effect, flament2017origin, cao2021coupled, flament2022assembly}.
This result illustrates how subduction history controls the lowermost mantle structure 
and suggests that the present-day dominant degree-2 lower mantle structure is not a stable feature throughout Earth's history \citep{Zhang2010, zhong2015temporal, muller2022tectonic} with both the number and location of hot structures likely having changed over time \citep{zhong2015temporal, flament2022assembly}. 
While this topic is still debated \citep{zhong2016long}, with some studies advocating for the long-term stability of the present-day degree-2 configuration \citep{bull2014effect, cao2021coupled}, there is agreement that the lowermost mantle structure strongly depends on plate motion history. 

Prior studies have defined criteria that mantle convection models used to provide heat flux patterns for geodynamo simulations should fulfill \citep{olson2016mantle}: (1) a realistic equation of state, considering phase transitions, mantle heterogeneities and initial and boundary conditions for temperature; (2) sufficiently complex rheologies considering pressure-, temperature- and strain rate dependent viscosity and possibly compositional variability; and (3) upper surface velocity boundary conditions derived from plate reconstructions over sufficiently long time spans.
However, mantle convection models computing CMB heat flux patterns have either used simple equations of state in incompressible models---neglecting important contributions to mass and energy transport and violating point (1), or they have studied CMB heat flux patterns arising in 
convection models without prescribed plate motions, so that plate-like behavior can emerge self-consistently, but generally does not resemble plate motions in the Earth's past---violating point (3).
Material properties controlling the transport of heat out of the core that have been simplified in models with prescribed plate motions also include lowermost mantle viscosity and thermal conductivity.
In particular the presence of a weak post-perovskite phase at the base of the mantle has been shown to have a strong impact on both pattern and magnitude of CMB heat flux \cite[i.e.,][]{nakagawa2011effects}. 
The thermal conductivity strongly varies with both pressure and temperature in the Earth's mantle, significantly affecting heat flux patterns and the differences between surface and CMB heat flux \citep{tosi2013mantle}.
Specifically, both mantle compressibility and the depth-dependence of the thermal conductivity act to decrease the volume of subducted slabs in the lowermost mantle: The lowered conductivity near the Earth's surface leads to slower cooling and thinner subducting plates, and the increasing density with depth causes slabs to decrease in volume as they sink. The slab volume is therefore overestimated in models that do not take into account these effects.
Neither of these complexities has been included in global 3D convection models with prescribed plate motions \cite[e.g.,][]{mcnamara2005thermochemical, ZHANG2011205, bower2013lower, bull2014effect, zhong2015temporal, hassan2015provenance, flament2017origin, cao2021coupled, cao2021evolution, flament2022assembly, muller2022tectonic, macleod2023evolution, frasson2023impact}, which instead assume a constant diffusivity.  
Finally, plate reconstructions that cover the whole supercontinent cycle have only become available recently \citep{MERDITH2021103477}. Even though these models can suffer from large uncertainties related to ambiguous paleomagnetic data, especially in the Precambrian, and uncertain subduction zone locations before 150~Ma, they are opening up the opportunity to study the associated changes in CMB heat flux patterns over sufficiently long time spans.

We here present global 3D mantle convection simulations that fulfil the criteria given above. These compressible, multi-phase, thermo-chemical convection models apply a plate reconstruction throughout the last 1 billion years \citep{MERDITH2021103477} as surface boundary condition to constrain the spatial and temporal heat flux variations at the core-mantle boundary and how they are affected by the physical properties of the mantle.  
We discuss how our models can be used to prescribe this heat flux as a boundary condition for geodynamo simulations, and make this workflow freely available, together with our model outputs. 
Our work provides a tool for the geodynamo community to estimate the variability of the core-mantle boundary heat flux and to use these patterns in geodynamo simulations in the future.

\section{Methods}

We set up our mantle convection models using the geodynamic modeling software ASPECT \citep{kronbichler2012high, heister2017high,gassmoeller_particles,clevenger_stokes19}. 
Specifically, we solve the equations for compressible mantle convection, using an equation of state that is based on mineral physics data (see Section~\ref{sec:equation_of_state}) and an Earth-like rheology (see Section~\ref{sec:rheology}). 
Since the focus of our study is deformation in the Earth's mantle, we consider only viscous stresses, and we assume that the viscosity is isotropic and that we can neglect terms including the bulk viscosity \citep{STO01}.
This leads to the following equations for conservation of mass, momentum and energy:
\begin{align}
  -\nabla \cdot \left( 2\eta \dot{\varepsilon} \right) + \nabla p &= \rho \mathbf g,
  \label{eq:aspect-system-1}
  \\
  \nabla \cdot (\rho \mathbf u) &= 0,
  \label{eq:aspect-system-2}
  \\
  \rho C_p \left(
    \frac{\partial T}{\partial t} + \mathbf u \cdot \nabla T
  \right)
  - \nabla\cdot(k \nabla T)
  & = \rho Q + 2\eta \dot{\varepsilon} : \dot{\varepsilon} +\alpha T \mathbf u \cdot \nabla p
  \label{eq:aspect-system-3}
\end{align}

where $\mathbf u$ is the velocity, $\dot{\varepsilon}$ is the deviatoric strain rate, $p$ the pressure and $T$ the temperature. 
Additionally, $\eta$ is the viscosity, $\rho$ is the density, $\mathbf g$ is the gravity vector, 
$C_p$ is the specific heat capacity of the material, $k$ is the thermal conductivity, 
$Q$ is the intrinsic specific heat production, and $\alpha$ is the thermal expansion coefficient. 

\begin{table}[h]
    \centering
    \caption{\label{tab:parameters} Model parameters.}
    \begin{tabular}{l l} 
    \hline
        Parameter & Value \\
        \hline
        Density $\rho$ & computed with Perple\_X$^{a}$ \\
        Thermal expansivity $\alpha$ & computed with Perple\_X$^{a}$ \\
        Specific heat $C_p$ & computed with Perple\_X$^{a}$ \\
        Thermal conductivity $k$ & 4.7 W/m/K or p-T-dependent (Table~\ref{tab:models}, Figure~\ref{fig:1d_profiles}b)\\
        Gravitational acceleration $g$ &  9.81~m/s$^2$\\
        \hline
        Viscosity formulation &  \citet{steinberger2006models}, see Figure~\ref{fig:1d_profiles}c\\
        Minimum viscosity & $10^{20}$~Pa~s\\
        Maximum viscosity & $5 \times 10^{23}$~Pa~s\\
        \hline
        Top boundary temperature & 273~K\\
        CMB temperature & 3700~K\\
        Potential temperature of inital adiabat & 1613~K\\
        Radiogenic heat production $Q$ & $2.09 \times 10^{-12}$ W/kg \citep{korenaga2017pitfalls}\\ 
        \hline
        Inner radius & 3481~km\\
        Outer radius & 6371~km\\
        Number of particles per cell & 15--120\\
        Model evolution time & 1000~Myr \\
        \hline
        \multicolumn{2}{l}{$^{a}$based on \citet{stixrude2011thermodynamics}}
    \end{tabular}
\end{table}
        
Our model geometry is a 3D spherical shell encompassing the whole mantle, and we prescribe the velocity at the surface based on a plate reconstruction of the last billion years \cite[see Section~\ref{sec:boundary_conditions}]{MERDITH2021103477}. During the model runtime we compute the heat flux at the core-mantle boundary using an accurate consistent boundary flux method described in Section~\ref{sec:cbf}.
To constrain the uncertainties in spatial heat flux variations, we compute several simulations with different material properties (Table~\ref{tab:models}), as described below.

\begin{table}
  \caption{List of model variations.}
  \label{tab:models}
  \centering
  \begin{tabular}{ll llll}
    \hline
    No. & Model & Basal layer & Thermal conductivity & Viscosity  \\
    \hline
    1 & Thermochemical  & basaltic & 4.7~W/m/K & reference$^{a}$ \\
    2 & Thermal         & none     & 4.7~W/m/K & reference$^{a}$ \\
    3 & p-T-dependent conductivity  & basaltic & p-T-dependent$^{b}$ & reference$^{a}$ \\
    4 & Weak post-perovskite  & basaltic & p-T-dependent$^{b}$ & ppv 100$\times$ weaker \\
    5 & Strong basalt   & basaltic & p-T-dependent$^{b}$ & basalt 10$\times$ stronger \\
    \hline
    \multicolumn{4}{l}{$^{a}$\citet{steinberger2006models}}
    \\
    \multicolumn{4}{l}{$^{b}$\citet{tosi2013mantle}, \citet{stackhouse2015first}}
  \end{tabular}
\end{table}

\subsection{Equation of State}
\label{sec:equation_of_state}

\begin{figure}
    \centering
    \includegraphics[width=\columnwidth]{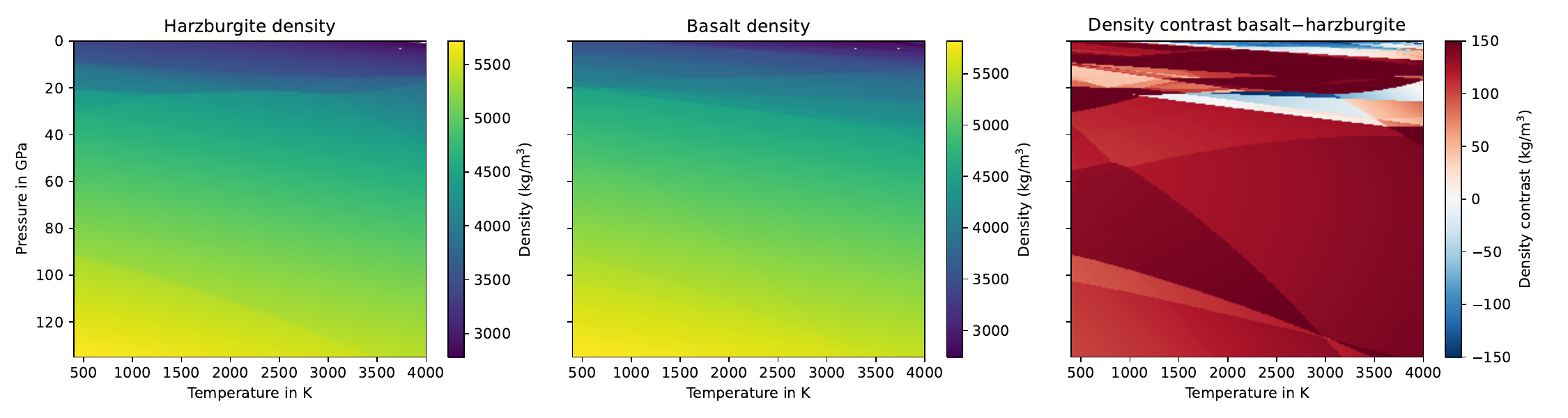}
          \caption{Density in dependence of temperature and pressure for harzburgite (left), MORB (center), and the density difference between the two compositions (right).}
          \label{fig:density}
\end{figure}

We treat the mantle as a mechanical mixture of mid-ocean ridge basalt (MORB) and harzburgite, using the composition from \citet{xu2008effect}. To compute the density, thermal expansivity and specific heat of the different lithologies, we use the thermodynamic modeling software Perple\_X \citep{connolly2009geodynamic} together with a thermodynamic database \citep{stixrude2011thermodynamics}. 
This approach automatically includes the effect of both compositional variations and mineral phase transitions on buoyancy, heat transport and volume changes/compressibility as described in previous geodynamic modeling studies~\citep{nakagawa2009incorporating}. 
The resulting density for basalt and harzburgite and the density difference between the two is shown in Figure~\ref{fig:density}. 
To compute the material properties of the mechanical mixture, we arithmetically average the material properties (based on a composition's volume fraction for density and thermal expansivity, and based on mass fractions for the specific heat).

\begin{figure}
    \centering
    \includegraphics[width=\columnwidth]{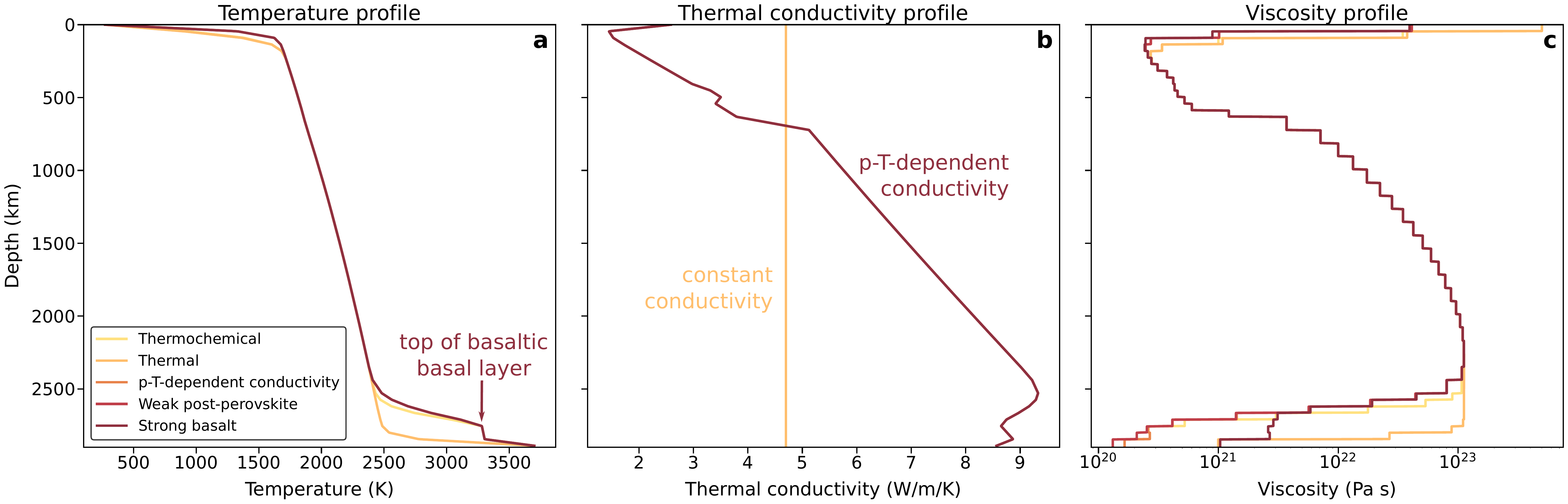}
    \caption{Profiles of temperature (a), thermal conductivity (b) and viscosity (c) at the start of each of the 5 different models listed in Table~\ref{tab:models}.}
    \label{fig:1d_profiles}
\end{figure}

We use a constant thermal conductivity of 4.7~W/m/K in our most simple setup. Since the thermal conductivity is expected to be much larger near the core-mantle boundary---increasing the CMB heat flux---we additionally investigate the effect of a pressure- and temperature-dependent formulation (see Table~\ref{tab:models}). Specifically, we use the thermal conductivity model from \citet{tosi2013mantle} above 660~km depth and the model from \citet{stackhouse2015first} below (Figure~\ref{fig:1d_profiles}b).

\subsection{Rheology}
\label{sec:rheology}

Our mantle rheology is depth- and temperature-dependent using the preferred viscosity profile of \citet{steinberger2006models} and an Arrhenius law that separates radial and lateral viscosity variations:
\begin{align}
  \eta(r, T) = \eta_0(r) \exp{ \left(-\frac{H(r) \Delta T}{n R T_\text{ref} T}\right)}
\end{align}
Here, $\eta_0(r)$ is the viscosity profile describing the depth-dependence, and $H(r)$ is the depth-dependent activation enthalpy \cite[as given in][]{steinberger2006models} defining the dependence on temperature. $T_\text{ref}$ is the reference temperature profile (the initial mantle adiabat), $\Delta T = T-T_\text{ref}$ is the deviation from this adiabat, $n$ is the stress exponent, and $R=8.314$~J/(mol~K) is the gas constant.
Since the focus of our study is the lowermost mantle, we do not take into account the strain rate dependence of the viscosity, but the stress exponent is assumed to be $n=3.5$ in the upper mantle and $n=1$ in the lower mantle to account for the change in dominant creep mechanism from dislocation to diffusion creep. 
We limit the viscosity to be between $10^{20}$ and $5\times10^{23}$~Pa~s. In particular the lower limit is chosen to make sure that convective processes do not occur on a smaller length scale than can be resolved by our numerical resolution. 

We also include two model setups that test additional rheologic complexities (see Table~\ref{tab:models}). The first one includes a viscosity reduction within the post-perovskite phase. Since the amount of weakening remains uncertain with estimates ranging from 1 to 4 orders of magnitude \citep{hunt2009weakening, ammann2010first, goryaeva2016low}, we here choose a viscosity reduction of a factor of 100. 
The other setup includes a composition-dependence of viscosity, increasing it by a factor of 10 in the basaltic material compared to the average mantle. This is motivated by the good fit to present-day mantle structure inferred from seismic tomography achieved in the models by \citet{flament2022assembly} using this value.
All viscosity profiles are shown in Figure~\ref{fig:1d_profiles}c. 

Since our models do not include a visco-plastic rheology or any other mechanism that would weaken plate boundaries, the prescribed plate motions at the surface (Section~\ref{sec:boundary_conditions}) lead to an unrealistically large amount of friction at plate boundaries. This is in particular the case in subduction zones, where the material is cold and viscous, and causes shear heating to be overpredicted. 
We therefore limit shear heating in Equation~\eqref{eq:aspect-system-3} in our models to prevent unrealistically high temperatures associated with the forced surface plate motion. 
Specifically, we compute if material would deform plastically as estimated by a Drucker--Prager yield criterion with a cohesion of $C=10$~MPa and a friction angle of $\phi = 0.085$. We then limit the stress being used to compute the shear heating to not be higher than this yield strength $\sigma_\text{yield}$.
\begin{align}
 Q_\text{shear} &= 
 \min(2 \eta \dot{\varepsilon}, \sigma_\text{yield}) : \dot{\varepsilon} \quad \text{with} \\
 \sigma_\text{yield} &= \frac{6 C \cos \phi + 6 p \sin \phi}{ \sqrt{3}  (3 + \sin \phi)}
  \label{eq:shear_heating}
\end{align}
This ensures that average mantle temperatures evolve as expected for the Earth's interior in our reference (No. 1; \textit{Thermochemical}) model, slowly cooling down over time. 

\subsection{Boundary conditions}
\label{sec:boundary_conditions}

To model the changing patterns of subduction throughout the supercontinent cycle, we use a reconstruction of the last 1 billion years of plate motion history \citep{MERDITH2021103477} and prescribe it as velocity boundary condition at the model surface. 
As the subducted slabs sink downwards into the lowermost mantle, they form regions of low temperature causing a high heat flux out of the core. In addition, they push hotter and/or chemically distinct material together into plume clusters or thermochemical piles, which feature low heat flux out of the core. This mechanism provides a coupling between the plate motions at the Earth's surface and the heat flux at the core-mantle boundary. Since the modeled subduction zones are prescribed in the same locations they are thought to have been on Earth, we also expect the pattern of core-mantle boundary heat flux to be representative of Earth's history.  
We realize that there are significant uncertainties associated with plate reconstructions, especially going further back in time than the oldest ocean floor preserved at present-day. 
Consequently, some subduction zones might be missing or be in the wrong location in our models. 
We nevertheless think that the use of this reconstruction is justified since the objective of our study is to predict characteristic changes in heat flux patterns throughout the supercontinent cycle rather than to constrain the exact heat flux pattern at the Earth's core-mantle boundary throughout the last billion years. 

The bottom boundary of our model is closed, but allows for free slip in the direction tangential to the boundary. 
The temperature is fixed to 273~K at the top and 3700~K at the bottom boundary. 

This core-mantle boundary temperature is well within the range of recent estimates. It can not be higher than the pyrolite solidus, which experimental data constrain to 3570$\pm$200~K \citep{nomura2014low}, 3430$\pm$130~K \citep{kim2020low}, or $\approx$3950~K with an uncertainty of 200--300~K \citep{pierru2022solidus}; and recent data on the melting curve of iron yield an estimate of 3760$\pm$290~K \citep{sinmyo2019melting}. 
Our value of 3700~K leads to a jump of approximately 1200~K across the thermal boundary layer. Uncertainties in the exact value of the CMB temperature should predominanly affect the average CMB heat flux and should only have a minor effect on the pattern of heat flux variations.

\subsection{Initial conditions}

Since the thermal and chemical state of the Earth a billion years ago is unknown, we here make the simplest assumption and start with an ``empty'' mantle (without any plumes or slabs) that is a mechanical mixture of 18\% MORB and 82\% harzburgite. Specifically, we assume that the initial temperature profile is adiabatic with a potential temperature of 1613~K with additional boundary layers at the top and bottom following a half-space cooling model.
The top thermal boundary layer has an age of 70~Myr, resulting in an appropriate temperature profile for oceanic plates.
The structure of the bottom boundary layer is different between our models setups (see Table~\ref{tab:models}). In the model without chemical heterogeneities (No. 2; \textit{Thermal} model), the bottom boundary layer is assumed to have an age of 50~Myr since the low viscosity allows for the frequent rise of plumes, keeping the layer relatively thin. 
In all other models, we consider the effect of a purely basaltic layer at the base of the mantle with an initial thickness of 200~km. Since this layer is dense, insulating the mantle and delaying plume formation, the thermal boundary layer is expected to be thicker compared to the purely \textit{Thermal} model. We therefore set the initial temperature to be 800~K above the mantle adiabat in the lowermost 150~km of the mantle (approximately 3300~K), 
and compute the temperature above that depth based on a half-space cooling model with an age of 150~Myr. The resulting temperature profile is shown in Figure~\ref{fig:1d_profiles}a. Note that since the thickness of the boundary layers is defined by age, it is different in the models with a pressure- and temperature-dependent thermal conductivity compared to the ones with a constant conductivity.

\begin{figure}
     \centering
     \includegraphics[width=0.925\textwidth]{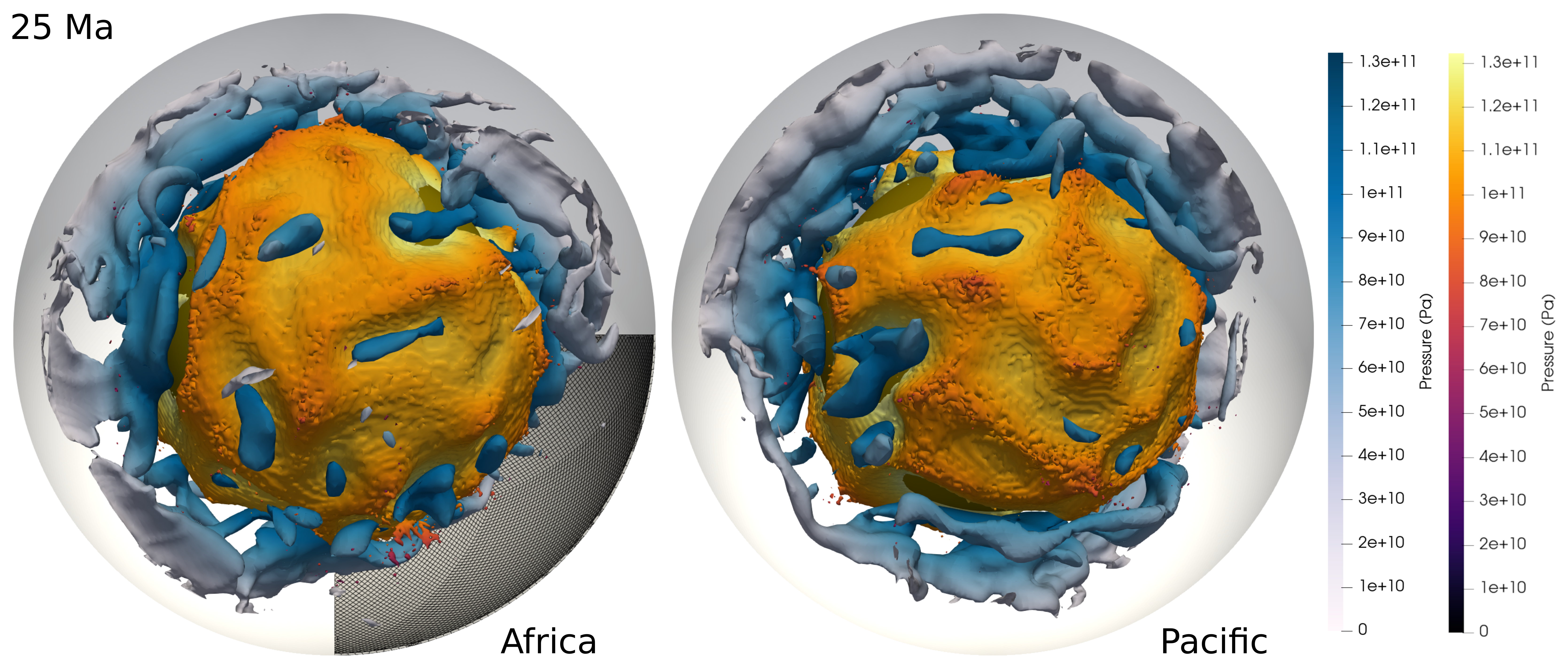}
     \caption{Visual representation of the mantle convection model, with slabs (regions at least $200$~K below the adiabatic temperature, blue-to-white colors indicating pressure, only shown below 670~km depth) and thermochemical piles (regions with a basalt fraction $>0.5$, yellow-to-red colors indicating pressure) interacting. The African hemisphere is shown on the left and the Pacific hemisphere is shown on the right. The outer model boundary is indicated by the gray sphere, the inner boundary (where not covered by slabs or thermochemical piles) is represented by the dark yellow sphere, with the computational mesh shown in the bottom right quadrant.}
     \label{fig:global_setup}
\end{figure}

\subsection{Numerical Methods}

ASPECT uses an adaptive finite-element mesh to discretize the model domain, which results in a cell size in our models between 45 and 250~km (depending on the refinement level and distance from the core-mantle boundary). Specifically, we refine the mesh in regions where the gradients in temperature, composition and viscosity are high, and we additionally enforce that the mesh is always refined to the highest resolution in the thermal boundary layers (below 2500~km depth and above 80~km depth). Since we use second-order finite elements, we achieve a resolution in terms of distance between the quadrature points of 39~km in horizontal and 22~km in vertical direction at the base of the mantle, where the core-mantle boundary heat flux is computed.

To minimize numerical diffusion, we use $\sim$90 million particles \citep[using the implementation of ][]{gassmoeller_particles} to track the evolution of the chemical composition. Specifically, our model contains two distinct compositions, harzburgite and mid-ocean ridge basalt. We assign each particle a composition that represents the fraction of basalt according to our initial conditions and interpolate particle properties to the finite-element mesh using a quadratic least-squares approximation.

\subsection{Computing heat flux with the consistent boundary flux method}
\label{sec:cbf}

The main purpose of our models is to compute the spatially and temporally variable heat flux density out of the liquid outer core and into the overlying mantle. However, accurately computing derived quantities like heat flux at the boundary of a numerical domain is challenging, because many numerical methods emphasize accuracy as an integrated quantity over the volume of a cell, and lose accuracy towards the faces and edges of a cell. In addition, the accuracy of a solution derivative is always reduced compared to the solution quantity itself. To circumvent these limitations, instead of computing the heat flux directly from the gradient of the temperature solution using Fourier's law, we utilize a consistent boundary flux (CBF) method as described for the heat equation in~\citet{gresho1987consistent}. CBF has been benchmarked in the geodynamics community as a very accurate method to compute another derived quantity---dynamic topography---and has also been suggested as a promising technique to compute heat flux before~\citep{zhong1993accurate, weir2019nodal}.

\begin{figure}
    \begin{center}
        \includegraphics[width=\columnwidth]{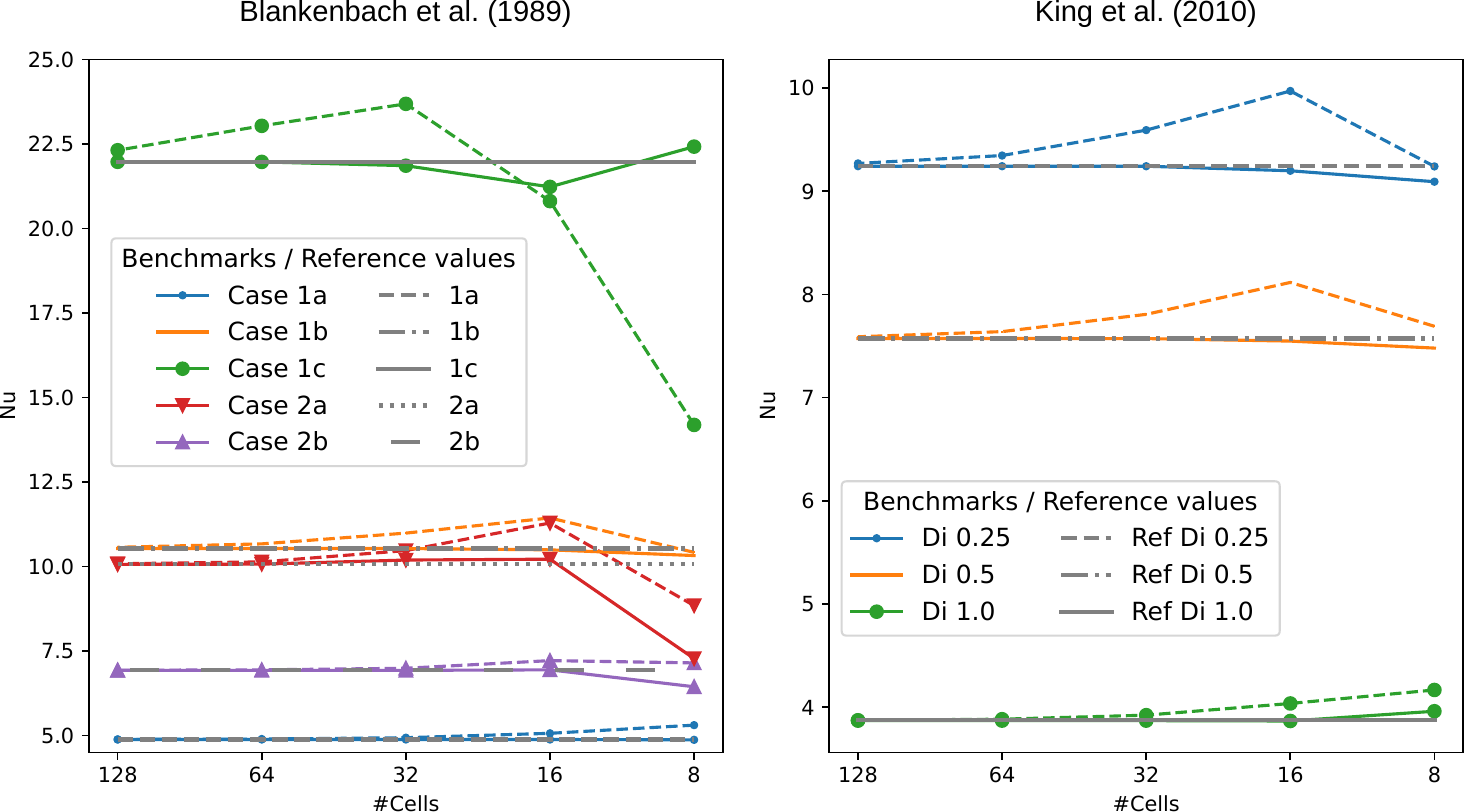}
        \caption{Nusselt number (boundary heat flux) over the number of cells (resolution) for the benchmark models of~\citet{blankenbach1989benchmark} (left panel) and~\citet{king2010community} (right panel). We plot results of the CBF heat flux method (solid lines) and a gradient-based heat flux method (dashed lines) compared to the reference results reported in the publications (gray lines). Note that the right panel only shows the case of $Ra=10^5$ and the anelastic liquid approximation for different dissipation numbers to improve the visibility of the figure and that we used the case ``RefVT'' as reference value.}
        \label{fig:blankenbach}
    \end{center}
\end{figure}

We have benchmarked our CBF implementation using the incompressible models described in~\citet{blankenbach1989benchmark} and the compressible models of~\citet{king2010community}, and illustrate the accuracy improvement of the CBF method in Figure~\ref{fig:blankenbach}. A description of the method and full benchmark results including a convergence analysis are provided in Appendix~\ref{appx:CBF} and all data to reproduce the benchmarks are included in ASPECT. In Figure~\ref{fig:blankenbach} we plot the heat flux postprocessing results for identical models computed with the CBF and traditional gradient based computations and compare them to benchmark results from the literature given above. It is very clear that while both methods converge to the reference results at high resolutions, the CBF method does so much faster and at much coarser resolutions (we show in the supplementary information that CBF's convergence order is 1-3 orders higher than the gradient based method). Additionally, the convergence of the CBF method is more consistent in the sense that it mostly approaches the reference value from one direction, while the gradient-based method consistently tends to underestimate heat flux at coarse resolutions and overestimate heat flux at intermediate resolutions. This behavior makes extrapolation of under-resolved model results---a common challenge in numerical geodynamics---much less accurate. Therefore, CBF heat flux represents a significant improvement in the accuracy of heat flux computations in geodynamic modeling studies.

\section{Results}

Our models compute the heat flux distribution across the CMB throughout the last 1 billion years of Earth's history. We will first compare the evolution of average properties (heat flux, temperature) and the present-day state of our models to available observations to show that our models are Earth-like. 
In a second step, we will analyze the computed heat flux patterns and constrain the possible spatial and temporal variations.

\subsection{Thermal evolution and average CMB heat flux}
\label{sec:thermal_evolution}

\begin{figure}
    \begin{center}
        \includegraphics[width=\columnwidth]{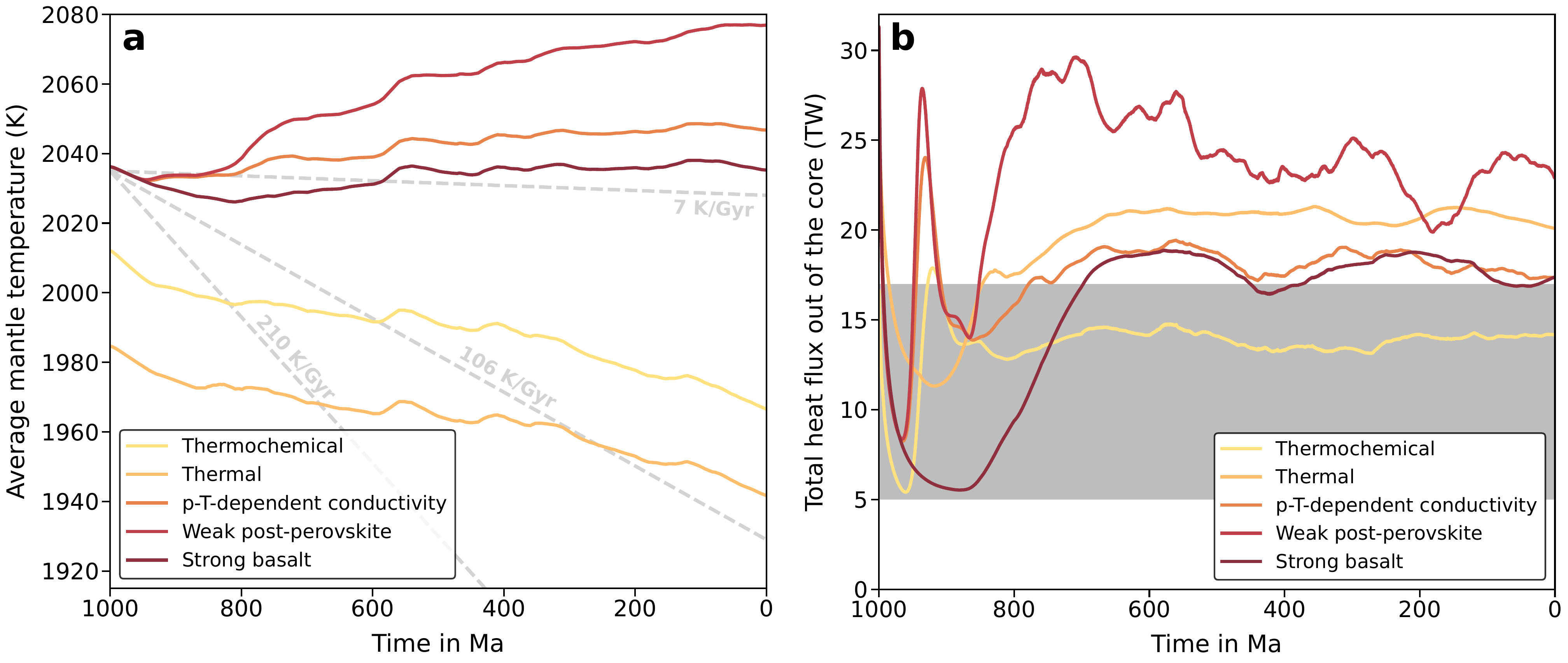}
        \caption{Change in average mantle temperature (a) and total heat flux out of the core (b) over time in all models. Estimates for the Earth's minimum, maximum and preferred cooling rate \citep[from][]{jaupart2015treatise} are marked as dashed gray lines for reference. Estimates for the Earth's CMB heat flux \cite[based on][]{jaupart2015treatise, nimmo2015energetics} are marked with a gray background.}
        \label{fig:temperature_over_time}
    \end{center}
\end{figure}

In our reference (\textit{Thermochemical}, Table~\ref{tab:models}) model setup, both the thermal evolution and the heat flux across the core-mantle boundary are consistent with observations.
The average temperature (Figure~\ref{fig:temperature_over_time}a) decreases over time in agreement with the cooling history of the Earth's mantle \cite[estimates for the present-day cooling rate are between 7 and 210 K Gyr$^{-1}$,][]{jaupart2015treatise}, and the CMB heat flux of 13--15~TW (Figure~\ref{fig:temperature_over_time}b) lies within the range of estimates \cite[5--17~TW, based on][]{jaupart2015treatise, nimmo2015energetics}. 
Based on the more recent, higher estimates of the core conductivity, heat fluxes near the upper end of this estimated range such as the ones in our models seem to be the most likely to be able to support a geodynamo. 

The choice of lowermost mantle properties strongly affects the computed thermal evolution. The thermal model without an intrinsically dense layer at the base of the mantle, which would insulate the core, exhibits a larger CMB heat flux. 
Incorporating a pressure- and temperature-dependent thermal conductivity reduces heat loss at the Earth's surface (due to the lower conductivity at low pressures, see Figure~\ref{fig:1d_profiles}) and increases heat flow across the CMB (due to the higher conductivity at high pressures), making mantle cooling less efficient.
The presence of a weak post-perovskite layer enhances convective heat transport away from the CMB, resulting in an increased heat flux. On the other hand, more rigid piles do not appear to substantially affect the amplitude of the CMB heat flow. 

In our computations, all models featuring a p-T-dependent conductivity---which is likely the better approximation
of Earth's mantle---display a thermal evolution inconsistent with observations (Figure~\ref{fig:temperature_over_time}a).
This suggests that our models underestimate the heat flux at the Earth's surface. One possible reason is that the data the plate reconstruction is based on becomes more sparse going back in time, and therefore some---especially intraoceanic---subduction zones may be missing in our boundary conditions. In addition, our models do not account for mechanisms such as melt extraction that facilitate more efficient heat transport across the lithosphere. 
However, since our focus is primarily on the general patterns of heat transport throughout the supercontinent cycle and their changes over time, small variations in average or upper mantle temperature do not substantially impact our results. Hence, we consider our models to be a reasonable approximation of the lowermost mantle processes under investigation.

Note that the heat flux density (Figure~\ref{fig:heat_flux_density}) goes through a phase of unrealistically strong variations in the first $\approx$200~Myrs of model evolution before featuring smaller variations around a quasi-steady-state. This effect is caused by the initialization of the models, which initially do not feature subducted slabs or rising plumes. Therefore, the heat flux first strongly decreases to lower values than expected on Earth as the core-mantle boundary region heats up, growing a thick thermal boundary layer that insulates the core. When the first cold slabs reach the CMB, they trigger the ascent of the first plumes and abruptly increase the amplitude of the heat flux. 
After this ``spin-up'' phase of approximately 200~Myrs, the modeled heat flux stabilizes and its variations reflect changes in the lower mantle structure due to the subduction history. We therefore do not interpret these first 200~Myrs of model evolution.

\subsection{Fit to present-day lowermost mantle structure}

Numerous studies have investigated the fit of lowermost mantle structure predicted from geodynamic models incorporating Earth's plate motion history to the LLSVPs observed in seismic tomography \citep{mcnamara2005thermochemical, bull2009synthetic, Zhang2010, davies2012reconciling, bower2013lower, flament2017origin, cao2021evolution, cao2021coupled, flament2022assembly, macleod2023evolution}, \cite[see][for a review]{mcnamara2019review}. 
All of these studies show that cold downwellings can push aside hot material residing at the base of the mantle---both in the form of dense thermochemical piles and in form of plume clusters---shaping it into a geometry roughly matching that of the LLSVPs. 
The fit between regions of high temperature in such a geodynamic model to low velocities in seismic tomography models can therefore serve as model validation. 

\begin{figure}
    \begin{center}
        \includegraphics[width=\columnwidth]{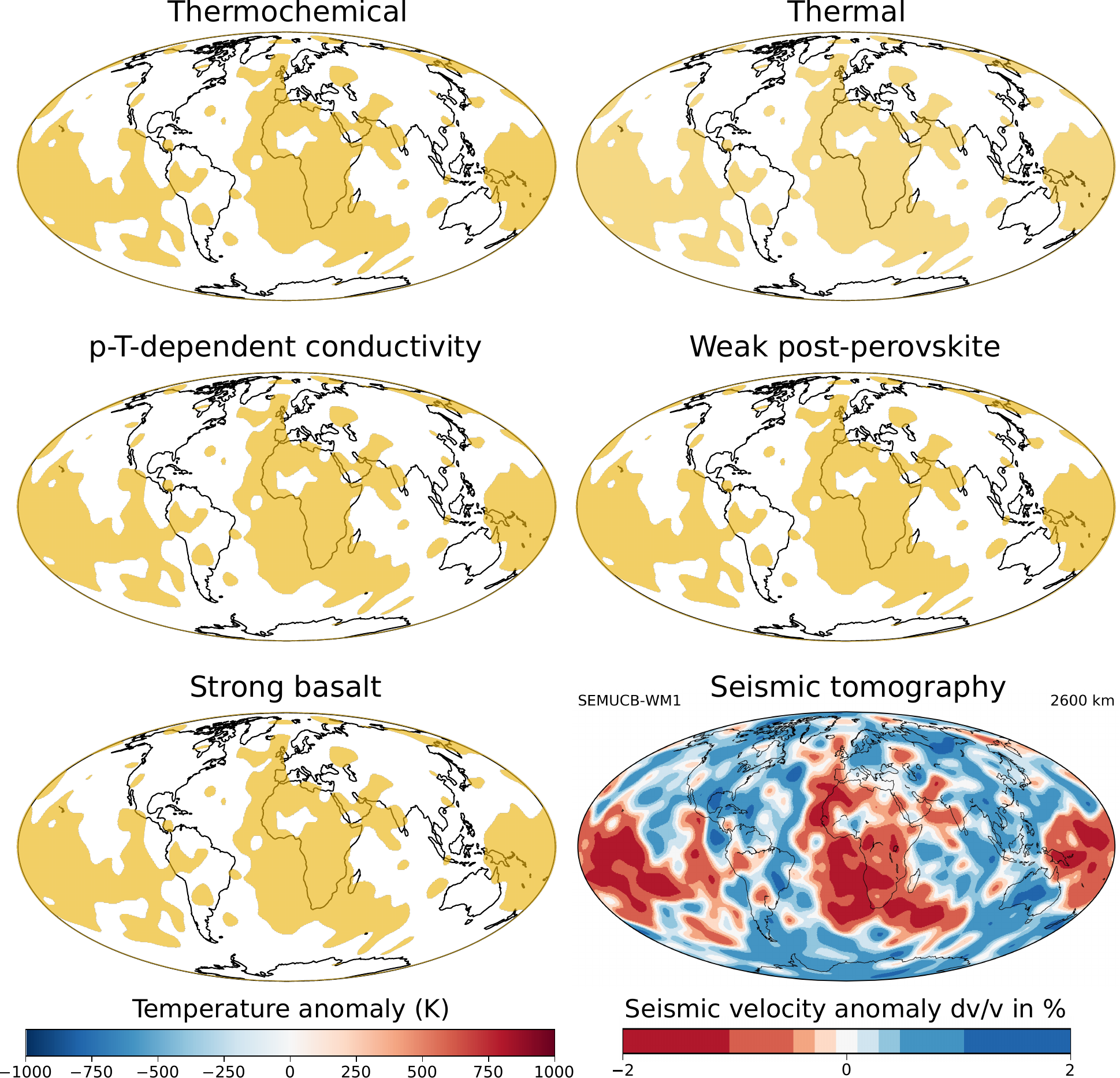}
        \caption{Temperature distribution (shown as the difference compared to the initial mantle adiabat) in a slice in 2600~km depth through all models at the end of the simulation (present-day). Bottom right panel shows the S-wave velocity anomaly of the seismic tomography model SEMUCB \citep{french2014whole} in the same depth slice for comparison. Yellow colors in the other panels indicate seismically slow regions (d$v/v < 0.1\%$) in SEMUCB.}
        \label{fig:temperature_2600km_depth}
    \end{center}
\end{figure}

We here present only a qualitative comparison, since our objective is not to to achieve the best possible fit of lowermost mantle structure to present-day Earth but to show that our models are characteristic for Earth's changing CMB heat flux patterns in general. In the \textit{Thermal} model, seismically slow regions are not matched well by regions of high temperature, with hot plumes being thin and roughly evenly spaced along the core-mantle boundary. However, all of our models that include an intrinsically dense layer show high-temperature anomalies at the base of the mantle in roughly the same regions as in the seismic tomography, i.e. below Africa and the Pacific (see Figure~\ref{fig:temperature_2600km_depth}). Similarly, areas of low temperatures are located below the ring of subduction surrounding the Pacific, both in the geodynamic models and seismic tomography. 
The exact geometry of the hot and cold regions varies between the different models and also differs from the shape of fast and slow anomalies in the tomography. Specifically, the hot regions in the geodynamic models are larger than the seismically slow regions in the tomography and in some models are broken up by colder patches that do not have seismically fast counterparts.  
However, the shape of the LLSVPs is reproduced fairly well in the \textit{p-T-dependent} model, with the African LLSVP being elongated in North--South direction and the Pacific LLSVP in East--West direction. 

We emphasize that a more quantitative comparison would require a conversion from temperature and composition to seismic velocities and applying a tomographic filter to make the amplitudes and gradients of the anomalies comparable. 
But even our more qualitative approach already shows that the match between geodynamic model and tomography is not as good as in some other recent models \cite[i.e.,][]{flament2017origin} whose plate motion history did not go as far back in time. 
This suggests that some of the older subduction zones ($>$250~Ma) that are less well-constrained in plate reconstructions deviate from their actual past locations in the Earth and that this old, cold material still affects the present-day lowermost mantle structure (i.e., visible as cold patches beneath Africa in the \textit{Thermochemical} and \textit{Strong basalt} models).  

Seismic tomography models also suggest that the LLSVPs are not flat features, but extend several hundred kilometers above the CMB (e.g., \citet{cottaar2016morphology} estimate the vertical extents of different parts of LLSVPs between 300~km and 1200~km). This behavior is not reproduced in our \textit{Thermal} model, but all models with an intrinsically dense layer feature high-temperature regions (${>750}$~K hotter than the mantle adiabat) at the base of the mantle that are at least 500~km high. The more effectively cold subducted slabs can displace this hot material, the more topography can be created, i.e., colder/denser/thicker slabs can generate taller piles of hot material. 
Therefore, the \textit{Thermochemical} model, which has a higher conductivity at the surface leading to thicker subducted slabs, features the tallest piles (600--800~km, compared to 500--700~km in the \textit{p-T-dependent} model).
In addition, weak thermochemical structures can not maintain a high topography, leading to a relatively low height of piles above the CMB (500--600~km) in the model with weak post-perovskite. 
However, the model with the lowest height of piles (400--500~km) is the one with a higher viscosity of basaltic material. Since convection within the strong piles is very sluggish, their tops cool down more efficiently than in the other models and hot regions do not extend as far upwards. 
Note that all the height values given represent the larger-scale structure of the piles, with individual plumes featuring high temperatures up to much shallower depth.
Given the uncertainties of seismic tomography models, any of the models with an intrinsically dense layer could be representative of the Earth's lowermost mantle structure.  

\begin{figure}
    \begin{center}
        \includegraphics[width=\columnwidth]{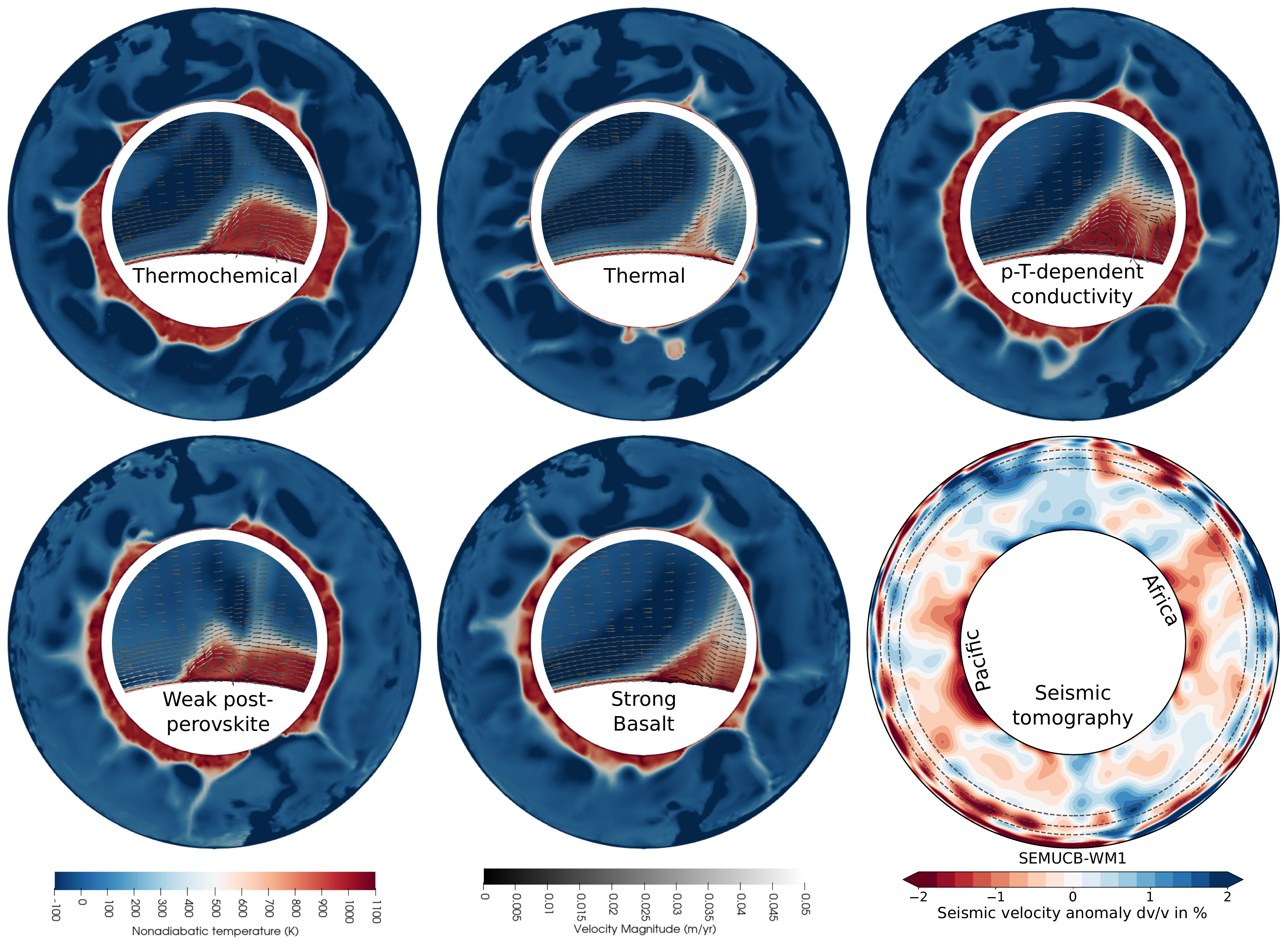}
        \caption{Temperature distribution (shown as the difference compared to the initial mantle adiabat) in an equatorial slice through all models at the end of the simulation (present-day). Insets in the center of each panel show a magnified version of one upwelling and one downwelling near the CMB and include arrows illustrating the flow field. The temperature scale is cut off at a negative temperature anomaly of 100~K to better show the internal thermal structure of the hot regions. For a version of the figure where the colorscale is symmetric about a temperature anomaly of zero instead, see Figure~\ref{fig:temperature_eq_slice2}. Bottom right panel shows the S-wave velocity anomaly of the seismic tomography model SEMUCB in the same equatorial slice for comparison.}
        \label{fig:temperature_eq_slice}
    \end{center}
\end{figure}

\subsection{Core-mantle boundary heat flux heterogeneity}
\label{sec:heterogeneity_amplitude}

While the total heat flux out of the core is important for driving the geodynamo, spatial heat flux variations along the CMB can affect the flow patterns in the core and the resulting magnetic field as well. 
Specifically, the amplitude of the heat flux heterogeneity and the heat flow near the equator have been identified as controlling factors \citep{glatzmaier1999role, kutzner2004simulated, olson2010geodynamo, olson2014magnetic}. 
In the following, we will quantify these spatial variations in terms of their amplitude, their characteristic length scales, and how both properties are expected to change throughout the supercontinent cycle. 

\begin{figure}
    \begin{center}
        \includegraphics[width=\textwidth]{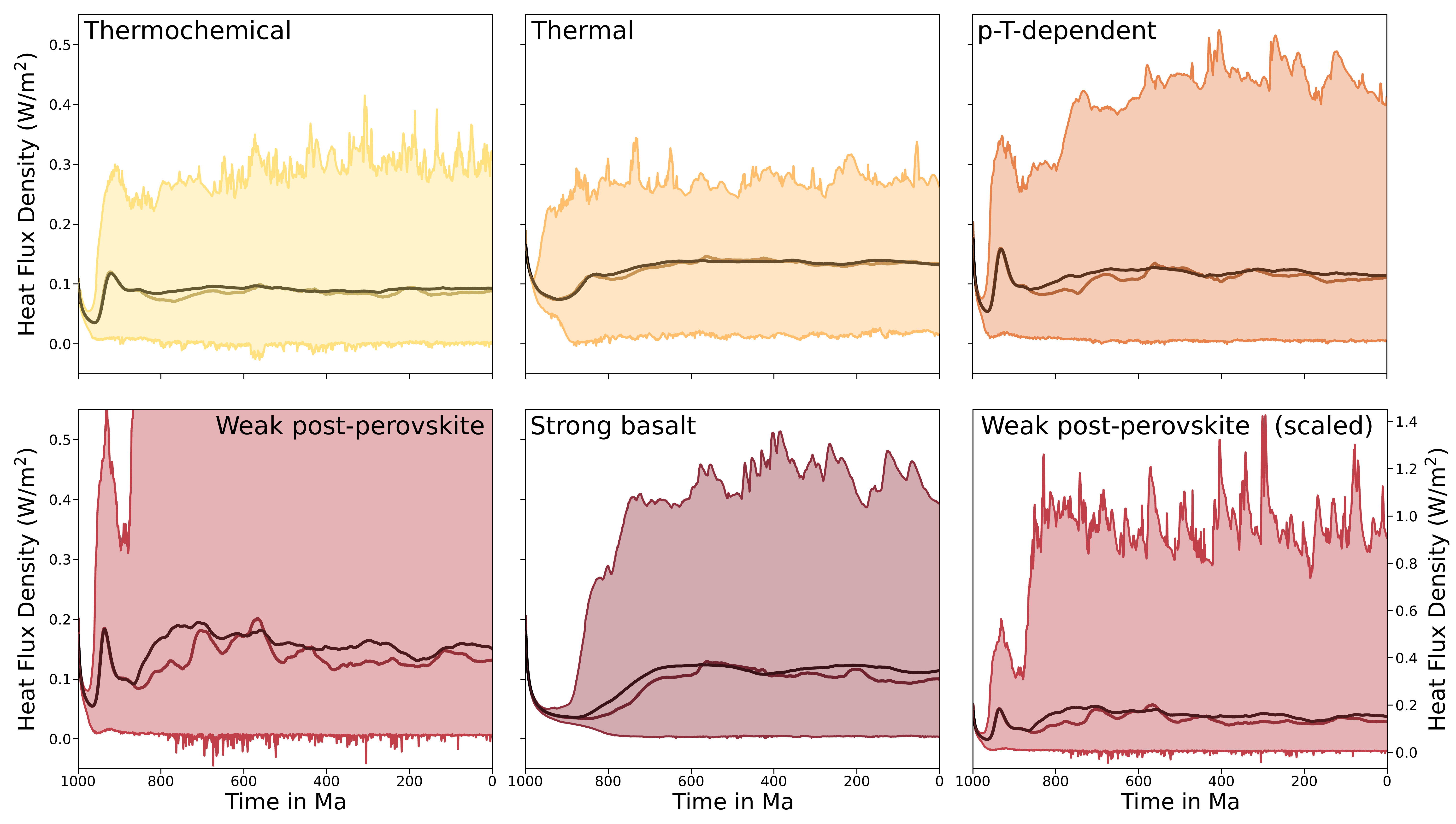}
        \caption{Changes in minimum, maximum, average (black line) and equatorial (darker colored thick line) heat flux at the core-mantle boundary over time for all models. The equatorial heat flux is defined as the average within $\pm 20\degree$ latitude of the equator \citep[as in][]{Zhang2010}. All panels use the same scale, except for the bottom right panel, which is re-scaled to show the full heat flux variations in the \textit{Weak post-perovskite} model.}
        \label{fig:heat_flux_density}
    \end{center}
\end{figure}

\begin{figure}
    \begin{center}
        \includegraphics[width=0.49\textwidth]{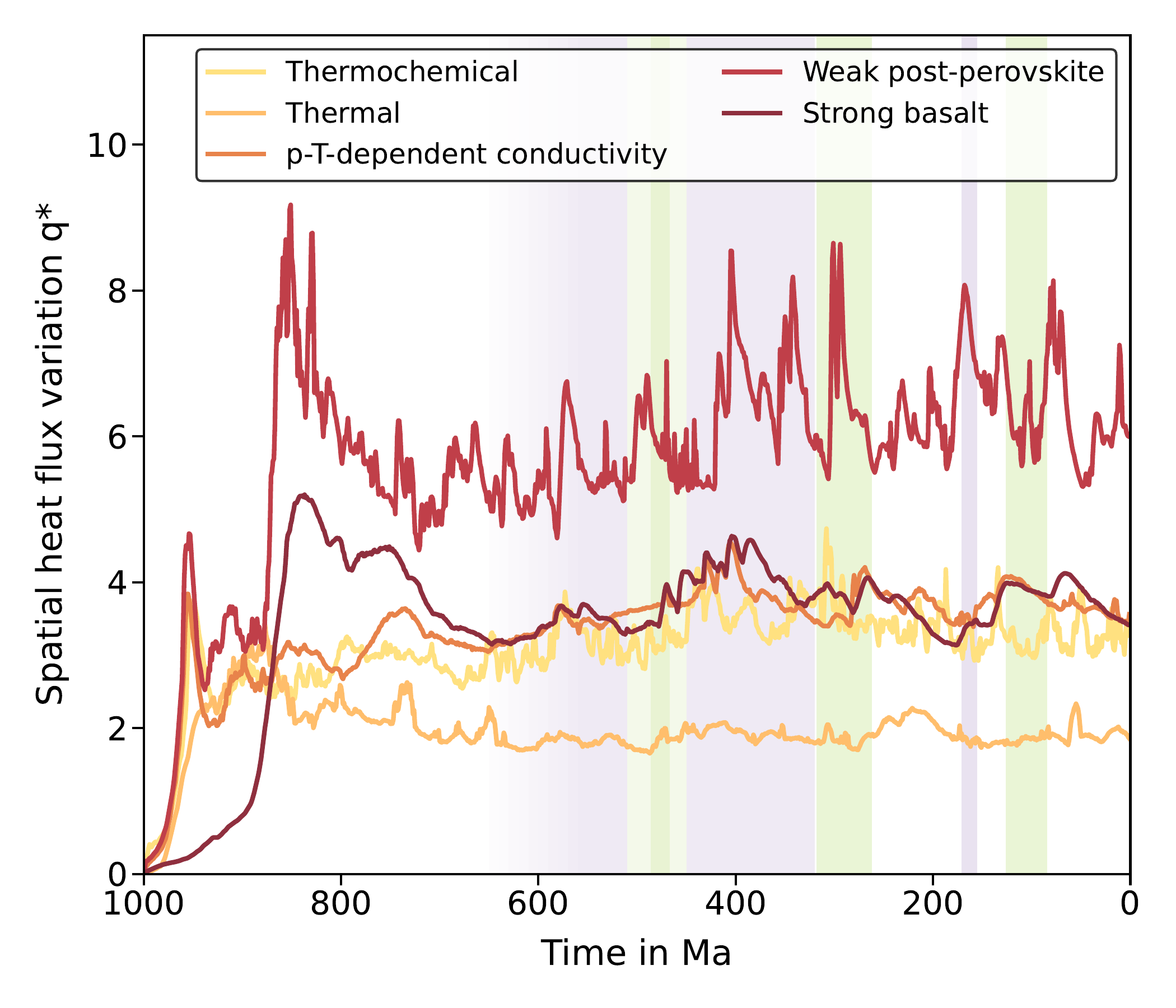}
        \includegraphics[width=0.48\textwidth]{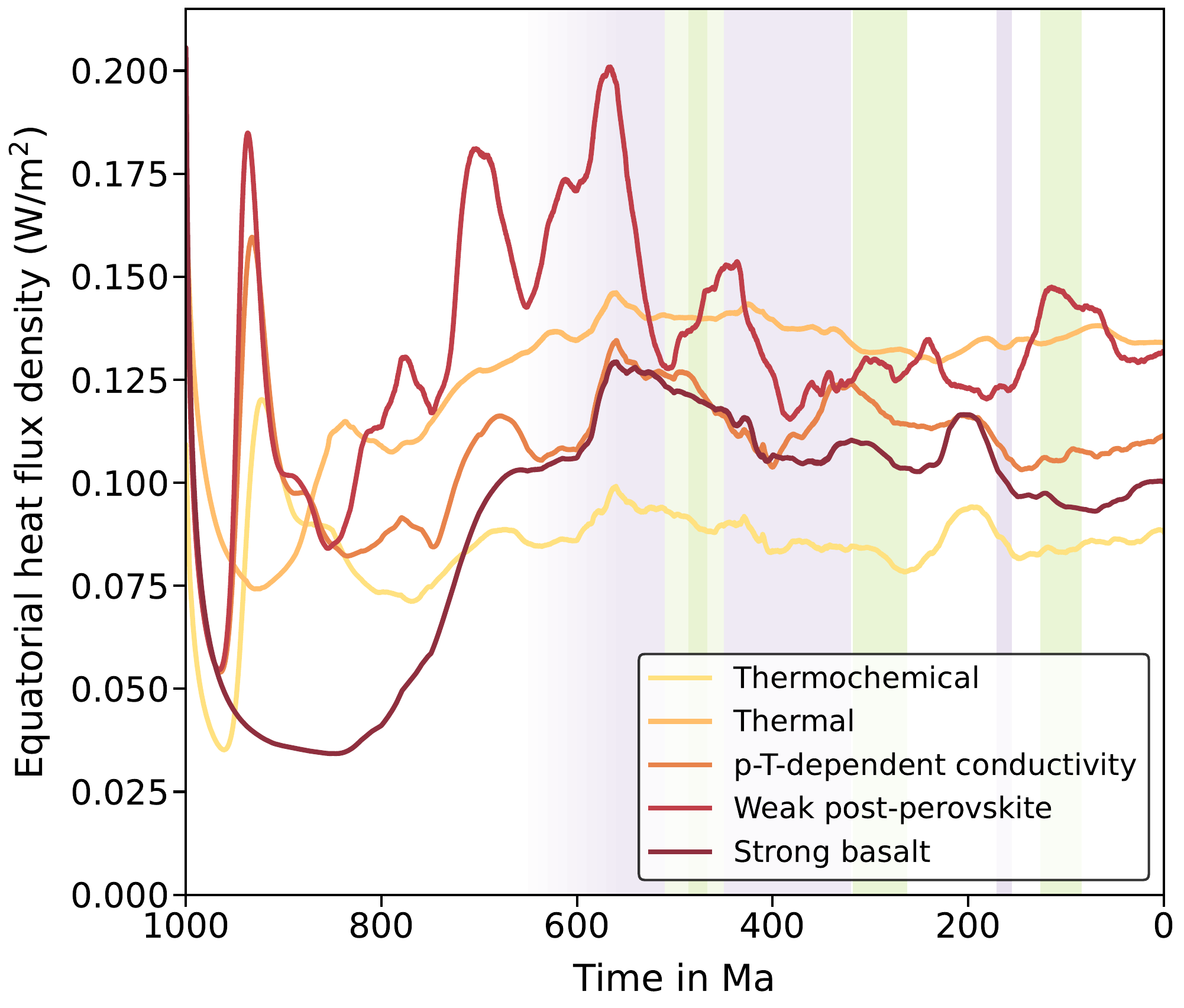}
        \caption{Evolution of q*, based on the minimum, maximum and mean heat flux shown in Figure~\ref{fig:heat_flux_density} (left), and the equatorial heat flux density (within $\pm 20\degree$ latitude of the equator, right). For reference, background colors show time periods of superchrons/low reversal frequency of the magnetic field (green) and periods of hyperactivity/high reversal frequency (purple), taken from \citet{hounslow2018subduction} for the last 500~Ma (darker shades) and from \citet{biggin2012possible} for suspected activity up to 600~Ma (lighter shades). Before 600~Ma, geomagnetic polarity is poorly known.}
        \label{fig:qstar}
    \end{center}
\end{figure}

When used as a boundary condition for geodynamo models, the amplitude of CMB heat-flux heterogeneity is often defined \cite[e.g.,][]{mound2017heat, sprain2019assessment, biggin2020quantitative} in terms of 

\begin{equation}
    q* = \frac{q_\text{max} - q_\text{min}}{q_\text{ave}}
\end{equation}
where $q_\text{max}$, $q_\text{min}$ and $q_\text{ave}$ are the maximum, minimum and spatially averaged heat flux across the CMB at a given point in time. We note that some studies follow a slightly different definition with an additional factor of 2 in the denominator \cite[i.e.,][]{olson2002time}, so care needs to be taken when comparing absolute values of $q*$.
Figure~\ref{fig:heat_flux_density} shows $q_\text{max}$, $q_\text{min}$ and $q_\text{ave}$, with the resulting $q*$ for all our models displayed in Figure~\ref{fig:qstar}. The minimum heat flux is close to zero in all models, representing areas at the base of the mantle with similar temperatures as the core and therefore a vanishing CMB heat flux. Conversely, the maximum heat flux varies over time much more substantially and is different between the different models. Note that our models include radiogenic and frictional heating so that locally, a small negative heat flux (out of the mantle into the core) is possible. 

In the \textit{Thermal} model, the minimum and maximum heat flux differ from the average by almost the same amount, with only weak variations in the maximum heat flux. 
This model also features a relatively low area fraction of the CMB with the heat flux being close to zero (i.e., yellow areas in Figure~\ref{fig:heat_flux_patterns}). Conversely, in all thermochemical models, the average heat flux is much closer to the minimum than to the maximum heat flux, with the latter also undergoing more substantial changes over time. The underlying cause is the low heat flux at the base of the hot and dense thermochemical structures, which cover a large fraction of the CMB area. The presence of these hot regions reduces the average CMB heat flux compared to an isochemical model that otherwise uses the same parameters, while the minimum and maximum heat flux remain almost unaffected.
The heat flux amplitude at the base of these piles is also different between the models. The more efficiently heat is both conducted and convected within the piles, the higher the CMB heat flux at their base. Consequently, the \textit{Strong basalt} model---where piles are harder to deform and convect more sluggishly---exhibits the lowest CMB heat flux within the piles (Figure~\ref{fig:heat_flux_patterns}). 
A pressure- and temperature-dependent conductivity---which increases the conductivity at the base of the mantle (see Figure~\ref{fig:1d_profiles})---amplifies the CMB heat flux both within piles and in regions where slab remnants accumulate compared to the \textit{Thermochemical} model (which has a constant conductivity). 
The \textit{p-T-dependent}, \textit{Weak post-perovskite} and \textit{Strong basalt} models therefore exhibit both a larger maximum and average heat flux. At the same time, the minimum heat flux remains near zero and does not undergo significant changes because the thermal gradient in the hottest regions remains low. 
The p-T-dependent conductivity also impacts the shape and distribution of hot piles and cold subducted material.
The lower conductivity within the lithosphere in these models leads to a thinner top thermal boundary layer, and therefore to thinner plates. After these thinner plates are subducted and sink down to the lowermost mantle, they also heat up more rapidly because of the increase in conductivity with pressure. 
Both effects lead to a decreased volume and a shorter preservation of cold anomalies at the CMB. 
Therefore, the fraction of CMB area covered by high-heat flux patches is larger in the \textit{Thermochemical} model---which features a constant conductivity---compared to the other models with chemical heterogeneities (Figure~\ref{fig:heat_flux_patterns}). 

The \textit{Weak post-perovskite} model exhibits the most extreme heat flux variations and the highest heat flux values out of all of the models (both locally and globally). This model features cold slabs that are substantially easier to deform compared to the other models, allowing cold material to spread out along the core-mantle boundary more quickly and facilitating the formation of localized cold ``puddles'' (see Figure~\ref{fig:temperature_eq_slice2}). Consequently, a strong thermal gradient develops, leading to a large heat flux out of the core. However, this increased heat conduction also causes rapid heating of the slabs, reducing the thermal gradient and resulting in a faster drop in the maximum heat flux. 
In the Earth, convection would be faster within the thermochemical piles as well, but due to our limit on the minimum viscosity (which is due to our model resolution) this effect is not included in our models. 

The different physical behavior of models with different material properties results in different characteristics of the amplitude of heat flux heterogeneities (as expressed by $q*$ in Figure~\ref{fig:qstar}). The \textit{Thermal} model features the lowest value, approximately 2, while the thermochemical models yield $q*$ values ranging from 3 to 4. The only exception is the \textit{Weak post-perovskite} model where $q*$ values range between 5 and 9. 
Note that these values usually need to be adjusted in geodynamo simulations by subtracting the adiabatic core heat flux, as discussed in Section~\ref{sec:application_to_geodynamo}.

The frequency of temporal variations is primarily influenced by the viscosity of the lowermost mantle. The \textit{Weak post-perovskite} model exhibits faster changes over time compared to the other models since material can deform more quickly. On the other hand, the heat flux changes more slowly in the \textit{Strong basalt} model since it takes longer to deform the dense piles at the base of the mantle.
Figure~\ref{fig:qstar} also shows periods of particularly high or low reversal frequency, since geodynamo studies suggest a connection to the amplitude of the CMB heat flux hererogeneity \citep[i.e.,][]{glatzmaier1999role, olson2010geodynamo, olson2014magnetic}.
However, we see no clear relationship between $q*$ and reversal frequency for any of our models.

Since the equatorial heat flux has been suggested to affect the magnetic field, and in particular its reversal frequency \citep[for example][]{olson2010geodynamo}, we have computed the averaged heat flux within $\pm 20\degree$ latitude from the equator (shown in Figure~\ref{fig:qstar}, and as thick colored lines partly overlapping with the average heat flux in Figure~\ref{fig:heat_flux_density}). 
In all models, the equatorial heat flux and the average heat flux follow similar trends, with both curves being particularly close in the \textit{Thermal} model (which features the lowest heat flux heterogeneity). 
When the equatorial heat flux deviates from the average, it generally exhibits lower values because the hot thermochemical structures are located close to the equator throughout most of the last billion years (see Figure~\ref{fig:heat_flux_patterns}). 
The different models do not follow the same evolution (Figure~\ref{fig:qstar}), highlighting the dependence of the timing of minima and maxima in both equatorial and average heat flux on lowermost mantle material properties. 
However, some trends are consistent in all models:
There is a maximum in the equatorial heat flux around 600 to 550 Ma, which is related to cold remnants of slabs subducted during Rodinia break-up. This cold material is advected along the CMB into the equatorial region from further South at $\sim$600~Ma due to the large-scale downwelling in the Southern hemisphere during Gondwana assembly. Another, smaller maximum can be seen in the \textit{Thermochemical}, \textit{p-T-dependent} and \textit{Strong basalt} models at $\sim$200~Ma. 
In addition, several models feature an equatorial heat flux minimum around 400 Ma, related to very little cold material being present near the equator (see Figure~\ref{fig:heat_flux_patterns}). 
Earlier modeling \citep{Zhang2010} has suggested that minima in the equatorial heat flux at 270 and 100~Ma may have been responsible for the Kiaman and Cretaceous Superchrons (which are marked as green bars in Figure~\ref{fig:qstar}). These minima are not consistently present in all of our models, with only the \textit{Strong basalt} model featuring low equatorial heat flux during both periods and none of the models showing a clear relation between equatorial heat flux and the observed reversal frequency.

\subsection{Temporal changes in core-mantle boundary heat flux patterns}

\begin{figure}
        \centering
        \includegraphics[width=\textwidth]{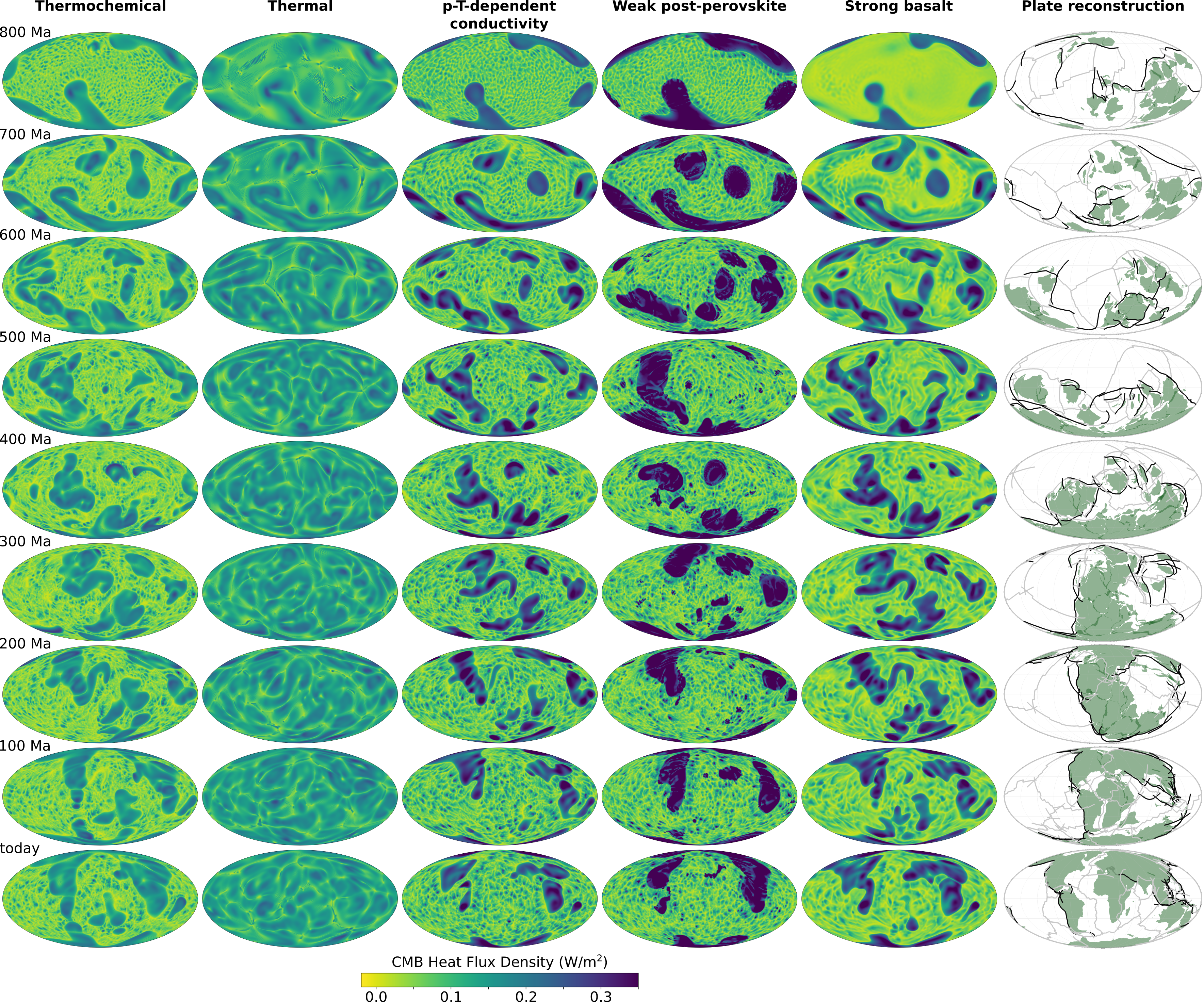}
        \caption{Evolution of the heat flux at the core-mantle boundary. Panels further up are further back in time. Large negative values (blue colors) indicate a large heat flux into the mantle and represent cold regions, values close to zero (yellow colors) indicate low heat flux into the mantle and represent hot regions, i.e. plumes or the hot thermochemical structures. The right column shows the location of plates (gray lines) and continents (green) in the plate reconstruction, with subduction zones marked in black. The complete evolution of all models is also displayed in Videos~S1 to S5.
        }
        \label{fig:heat_flux_patterns}
\end{figure}

While the amplitude of heat flux heterogeneities is likely to affect the geodynamo, the distribution of these heterogeneities could have an important effect as well. 
Figure~\ref{fig:heat_flux_patterns} shows the changes in these CMB heat flux patterns throughout the supercontinent cycle. 
All models show low CMB heat flux (yellowish colors) where hot material is located: Localized at the base of plumes in the \textit{Thermal} model; more spread out within the thermochemical piles in all models with chemical heterogeneities. Since these piles convect internally (highlighted by the honeycomb-like pattern within these regions) but are too dense to advect this heat further upwards, they maintain relatively high temperatures and insulate the core. 
Conversely, regions where subducted slabs reach the CMB and increase the local thermal gradient feature high heat flux (bluish colors). 
Variations between areas of high and low heat flux occur on several different characteristic spatial scales: 
In the \textit{Thermal} model, patches of high and low heat flux alternate on intermediate length scales ($\sim$1000--2000~km), with cold subducted material and hot upwellings distributed relatively evenly along the whole CMB. 
This is in contrast to the models with dense thermochemical structures, which occupy large areas between the subducted slabs, preventing them from spreading out across the whole CMB and leading to larger-scale variations in heat flux, with large high- and low-heat-flux patches located in specific parts of the globe. 
In addition to these global-scale heterogeneities, there is an additional, superimposed small-scale pattern within the piles reflecting their internal convection. The characteristic length scale of these convection cells is controlled by the viscosity within the piles, with a low viscosity (as in the \textit{Weak post-perovskite} model) leading to smaller-scale structures, and higher viscosity (such as in the \textit{Strong basalt} model) leading to larger-scale structures.

In all models, the heat flux patterns show characteristic changes throughout the supercontinent cycle, controlled by the subduction history in the plate reconstruction (Figure~\ref{fig:heat_flux_patterns}, right column).
Wherever subduction zones are located at the surface, cold material sinks downwards, reaching the CMB about 150--200~Myrs later and leading to an increased CMB heat flux. 
This becomes apparent when comparing the regions of high heat flux (blue colors in the first 5 columns in Figure~\ref{fig:heat_flux_patterns}) to the trench locations (black lines in the last column) 100--200 Myr earlier (one or two rows further up). 
For example, the girdle of high heat flux separating the two low heat flux piles between Africa and Pacific at the present-day (last row in Figure~\ref{fig:heat_flux_patterns}) corresponds to the ring of subduction zones surrounding Pangea as it was breaking apart (the black lines surrounding the continents in the last column and third-/second-to-last row in Figure~\ref{fig:heat_flux_patterns}). 
Conversely, during the time frame 600--500 Ma, the later stages of the Gondwana assembly, most of the subduction zones were located in one hemisphere (centered around where the Southern Atlantic Ocean is located today). This is reflected in the heat flux patterns, especially in the \textit{Thermochemical}, \textit{p-T-dependent} and \textit{Strong basalt} models at 400~Ma, which all feature one hemisphere with several high heat flux patches, while the other hemisphere consistently exhibits much lower heat flux. In particular the subduction between Australia--Antarctica and the Proto-Pacific Ocean at 600~Ma is clearly visible as a three-forked high heat flux patch between $\sim$60--120$\degree$ W (approximately across the area where North and South America are located today) at 400~Ma. 
Instead of a hemispherical distribution, at earlier times around 600~Ma, patches of high heat flux are spread out across the whole globe. 
This pattern matches the trench locations at 800--700~Ma, which are similarly distributed, with additionally several subduction zones forming, disappearing or changing location in the transition from Rodinia break-up to Gondwana assembly. 
Finally, the girdle of high heat flux surrounding the single thermochemical pile in the 800~Ma panel in the models with chemical heterogeneities corresponds to the ring of subduction zones surrounding Rodinia $\sim$150--50~Myr prior. 
How long exactly it takes for material subducted at the surface to affect the CMB heat flux, and for how long the resulting cold anomalies are preserved at the CMB depends on the material properties of the individual model. For example, in the \textit{Weak post-perovskite} model the cold remnants of subducted slabs are deformed more easily, heat up faster, and impact the CMB heat flux for a shorter time, whereas convection at the CMB is more sluggish in the \textit{Strong basalt} model so that high-CMB heat flux patches are preserved for a longer time.

\begin{figure}
        \centering
        \includegraphics[width=.95\textwidth]{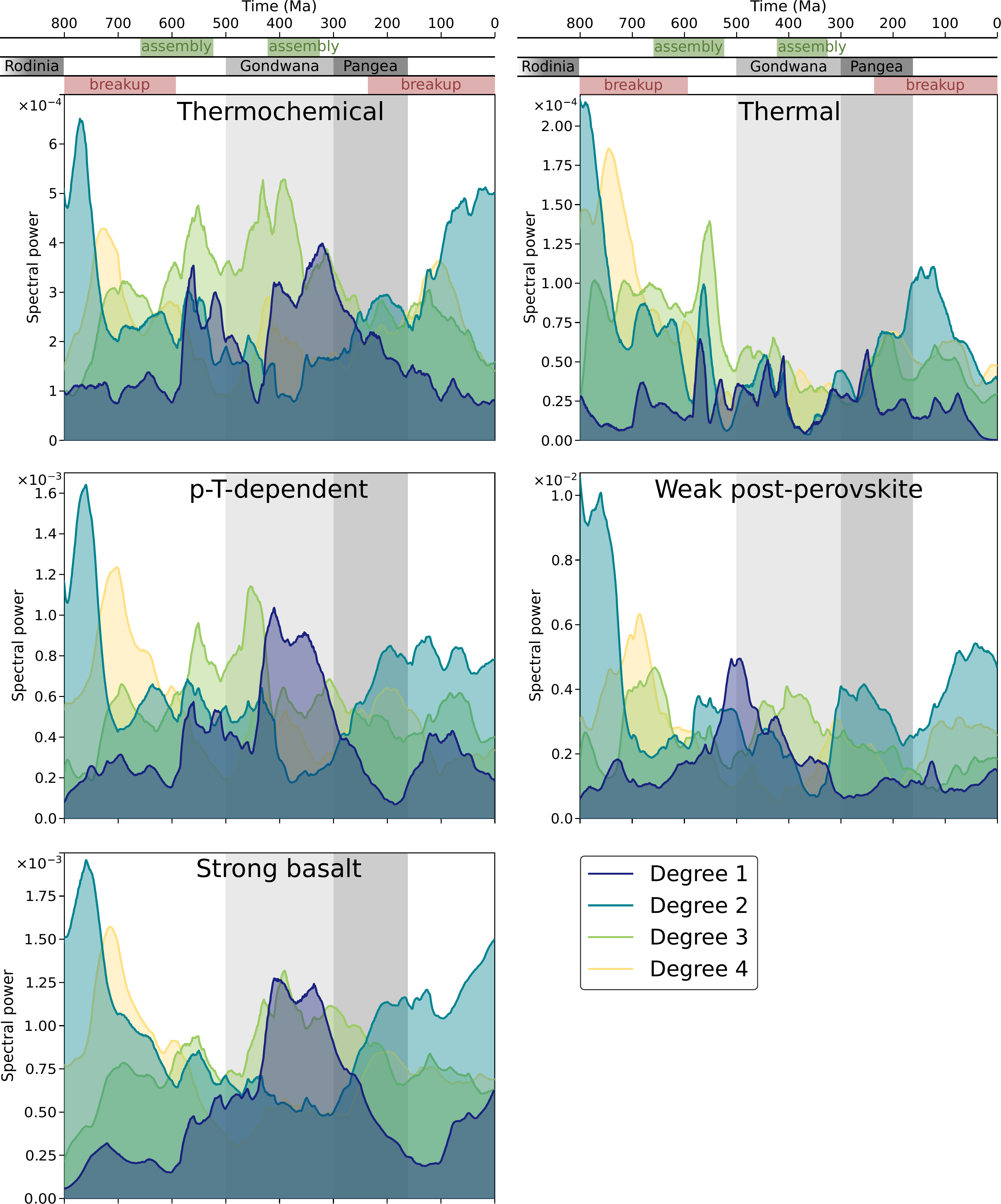}
        \caption{Changing power of the first 4 spherical harmonics degrees of the core-mantle boundary heat flux over time. The scale at the top indicates the supercontinent cycle \citep[times taken from][]{merdith2019rift}.}
        \label{fig:SH_degrees}
\end{figure}

This connection between the supercontinent cycle and CMB heat flux patterns becomes even clearer in a more quantitative analysis of the temporal variations of their characteristic wavelengths using spherical harmonics. 
To evaluate which spatial pattern is most prevalent, we assess the power spectrum of the different spherical harmonics degrees over time (Figure~\ref{fig:SH_degrees}). 
Degree 1 dominating the spectrum indicates a large-scale heat flux difference between the two hemispheres, one featuring high heat flux, the other featuring low heat flux.
Conversely, a high power in degree two corresponds to two large regions of low heat flux separated by a band of low heat flux (or the other way around). If higher degrees are dominant, spatial variations on a smaller scale are more prevalent. 

Figure~\ref{fig:SH_degrees} shows how the prevalence of the different degrees changes throughout the supercontinent cycle. 
At around 800~Ma, and then again around 200--0~Ma, there is a high power in degree 2 in all models, corresponding to two high-temperature (low heat flux) structures in the lowermost mantle that are split by a band or several patches of cold remnants of subducted slabs (high heat flux; see Figure~\ref{fig:heat_flux_patterns}). In both instances, this pattern succeeds the start of supercontinent (Rodinia or Pangea) break-up, where subduction zones surround the supercontinent. As this girdle of cold subducted slabs sinks down to the lowermost mantle, it displaces hot material, splitting it into two structures, one beneath the (disassembling) supercontinent, one beneath the superocean. This is also the stage that we observe for the present-day Earth. 
The later stages of the break-up and the transition to the assembly of the next supercontinent requires subduction zones to be more spread out. The \citet{MERDITH2021103477} plate reconstruction also features a rapid reorganization of plate boundaries during this time frame (around 850--700~Ma). This is reflected in a low power of both degree 1 and 2 as this subducted material reaches the lowermost mantle (around 700--600~Ma), with higher spherical harmonics degrees being more dominant and the corresponding smaller-scale alternation between hot and cold material along the CMB. 
The only time where degree 1 is dominant in some of our models (and the power of degree 2 is low across all models) is following the assembly of a supercontinent (i.e., around 450 to 300~Ma). Because the subduction zones are distributed predominantly in the Southern hemisphere in \citet{MERDITH2021103477} at 600--450~Ma, where Gondwana is being assembled, cold material also reaches the lowermost mantle only in one hemisphere. This leads two one hemisphere with predominantly cold, subducted material (high heat flux) below the supercontinent and one hemisphere with predominantly hot material (low heat flux) below the superocean. Due to the delay of 150--200~Myrs between subduction zones in the plate reconstruction being able to affect CMB heat flux, the prevalence of degree 1 then occurs after the assembly and during the supercontinent stage. 
Note that the subducted material can easily be pushed along the CMB once it has reached the lowermost mantle because of the negligible friction at the CMB. Because cold material is pushed northwards in our simulations (which is mostly an effect of a net rotation of the whole mantle between 450 and 400~Ma), the location of the high heat flux hemisphere (centered around present-day Africa) does not exactly correspond to the hemisphere where subduction occurred (centered around the South Pole). 

The alternation between a degree 1 and 2 lower mantle structure has been debated over the last decades. While some studies argue for the prevalence of a degree-1 structure before Pangea formation \citep{Zhang2010, zhong2015temporal} and the aggregation and dispersal of basal mantle structures over time \citep{flament2022assembly}, other studies have suggested that lower mantle structure is dominated by spherical harmonic degree 2 most of the time \citep{cao2021coupled} and that the two LLSVPs observed today have remained close to their present-day positions for at least the past 410 Myr \citep{bull2014effect}.
Despite all our models (other than the \textit{Thermal} model) showing this alternation between degree 1 and 2 throughout the supercontinent cycle \citep[in agreement with][]{Zhang2010, zhong2015temporal, flament2022assembly}, they still do not provide a conclusive answer to this debate. As we have shown, the temperature and heat flux distribution in the lowermost mantle are directly controlled by subduction history and therefore depend on the specific plate reconstruction being used. The plate reconstructions used in \citet{Zhang2010, bull2014effect, cao2021coupled, flament2022assembly} differ from each other, likely contributing to their contradicting results.
Going back in time, reconstructed features become more uncertain, and especially intra-oceanic subduction zones are very hard to constrain. 
While supercontinent break-up requires subduction zones around the supercontinent, and the assembly of a supercontinent requires subduction zones between the terranes being assembled, the subduction history in the superocean hemisphere is unclear. The presence of additional subduction zones in this hemisphere during Gonwana assembly could break the degree 1 pattern we see in our models around 450 to 200~Ma. 
Our models show that a cycle between degree 1 and 2 patterns is a plausible scenario, but better constraints on past subduction zone locations are required to determine with certainty if this cycle has occurred in Earth's history.

\section{Applying core-mantle boundary heat flux patterns to geodynamo simulations}
\label{sec:application_to_geodynamo}

We envision our computed CMB heat flux patterns to be used as boundary conditions for geodynamo simulations. 
Therefore, we provide a time series of heat flux patterns at the CMB for each of the presented global mantle convection models that can be used as boundary conditions for numerical geodynamo simulations. These datasets comprise a spherical harmonic description of CMB heat flux patterns for each mantle convection model in $\sim$1 Myr ($\pm$1.2~kyrs) time steps that were converted from the model output generated with ASPECT. For each time step, the CMB heat flux data calculated by the ASPECT models is given at 49,151 equidistant points on a spherical surface at radius $r = 3.481 \times 10^{6}$~m in W~m$^{-2}$, with the topology of the mesh being based on a decomposition of the sphere into six identical regions \citep[see, for example,][]{thieulot2018ghost}.
The 'outward' heat flux from the core into the mantle is defined as negative. 

To convert this heat flux to a spherical harmonics representation, as generally used in geodynamo codes, we interpolated it with PyGMT \citep{pygmt2023} onto a regularly spaced grid that accommodates the \citet{DriscollHealy1994} sampling theorem and is of the size $N\times 2N$, with $N = 2 L_\text{max} + 2$. Using an $L_\text{max}$ of 256, this results in a quadrature point spacing of $\Delta\theta = \Delta\phi = 0.35^{\circ}$. We then expanded this regularly gridded data to spherical harmonics as real Schmidt semi-normalized harmonics using the Python module pyshtools \citep{shtools2018}, with L\textsubscript{max} = 256, and the Condon-Shortley phase factor of $-1^{m}$ appended to the associated Legendre functions \citep{holmes2002unified}. For ease of use, we provide the spherical harmonic description of the CMB heat flux patterns for each model in NetCDF format as part of our accompanying data publication \citep{dannberg2023changes}. Note that because it takes the cold slabs approximately 190~Myrs to reach the CMB after being subducted, the maps of the first $\sim$200~Myrs of the model runs do not represent realistic depictions of the CMB heat flow patterns. We nevertheless provide CMB heat flux maps for the complete time span of the model runs. 
To be used as realistic boundary conditions in numerical geodynamo simulations, the spherical harmonics can be scaled freely to account for the core's adiabatic heat flow as outlined below, and they might need to be converted to non-dimensional parameters depending on the specific dynamo model being used. 

Geodynamo simulations typically assume the Boussinesq approximation, which does not take into account adiabatic heating. 
The heat flux imposed at the core-mantle boundary therefore needs to be adjusted by subtracting the conductive heat flux along the core adiabat from the CMB heat flux provided by the mantle convection model. 
This adiabatic core heat flux depends on the thermal conductivity and the material properties that determine the temperature gradient along the core adiabat, so the value that needs to be subtracted will depend on the specific setup chosen for an individual geodynamo simulation. 
However, we here also want to demonstrate how this conversion would affect the amplitude of spatial heat flux variations as seen in the geodynamo model, i.e., the $q*$ parameter. 
We choose three different values of 5, 10, and 15~TW for the adiabatic core heat flux, within the range of recent estimates \cite[2.3 to 16 TW,][]{pozzo2012thermal, nimmo2015energetics, davies2015constraints, silber2019heat, mound2023longitudinal}. 
In addition, we assume that the $q_\text{max}$, $q_\text{min}$ and $q_\text{ave}$ cannot change sign from subtracting the adiabatic heat flux, i.e., heat will not flow out of the mantle into the core because the CMB heat flux computed in the mantle model has a smaller amplitude than the conductive heat flux along the core adiabat.

\begin{equation}
    q* = \frac{\max(q_\text{max}-q_\text{adi}, 0) - \max(q_\text{min}-q_\text{adi}, 0)}{\max(q_\text{ave}-q_\text{adi}, 0)}
\end{equation}
with $q_\text{adi}$ being the adiabatic core heat flux density. 
We ignore results where the average CMB heat flux is smaller than the adiabatic core heat flux since in these cases, heat loss to the overlying mantle would likely not be able to drive the geodynamo. 
%[TODO: If we want to do this properly, we would have to do this pointwise...]

\begin{figure}
        \centering
        \includegraphics[width=\textwidth]{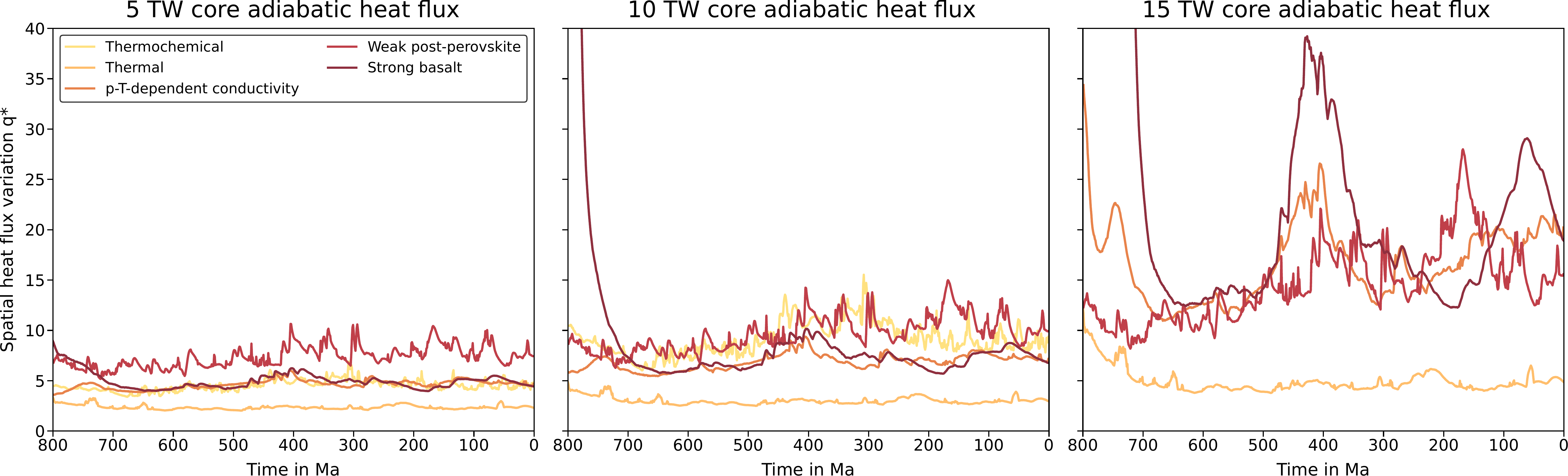}
        \caption{Evolution of $q*$ for all models when subtracting the the heat flux conducted along the outer core adiabat, assumed to be 5~TW (left), 10~TW (middle), 15~TW (right). Since 15~TW is larger than the average heat flux in the \textit{Thermochemical} model, this model would not have a net heat flux out of the core and is not shown in the right panel.}
        \label{fig:qstar_with_Qadi}
\end{figure}

The resulting values for $q*$ are shown in Figure~\ref{fig:qstar_with_Qadi}. 
In all cases (including the case where $q_\text{adi}$ is not subtracted), $q* \geq 2$, imposing a minimum on the expected amplitude of heat flux heterogeneity at the core-mantle boundary. Since estimates for the adiabatic core heat flux are similar to the estimates for the total heat flux out of the core, it is impossible to provide an upper limit to $q*$, and the case with the 15~TW adiabatic core heat flux illustrates that $q*$ can easily reach values $>30$. 
If we further assume that the adibatic core heat flux is at least 10~TW (based on the more recent higher estimates for the thermal conductivity of the core) and that there is a layer of dense material at the base of the mantle, then we would expect values of $q* \geq 6$. 

 %[Notes on $Q_a$ at CMB:
%15 to 16 TW \citep{pozzo2012thermal},
%2.3 to 15.2 TW \citep{mound2023longitudinal},
%15 TW \citep{nimmo2015energetics},
%4 – 17 TW %\citep{davies2015constraints},
%9.4–12 TW \citep{silber2019heat}]

%[Write about non-dimensionalization, conversion to spherical harmonics, data formats, and making our output available]

\section{Model limitations}

We have already outlined some of the limitations of our models above, such as the uncertainty in the plate reconstructions for earlier times in Earth's history and the relatively large total CMB heat flux in some of our models. Another model simplification is that our CMB temperature remains constant over time rather than evolving based on the amount of heat extracted from the core. While these factors impact the evolution of the total CMB heat flux and the manifestation of specific CMB heat flux patterns, we do not expect them to significantly affect the amplitude of spatial and temporal CMB heat flux variations, which is the focus of our study.
Below, we discuss some additional factors that are important for understanding how our results can be applied to gain insights about outer core convection. 

While our models include a pressure- and temperature-dependence of the thermal conductivity, they do not explicitly incorporate how mineral phase changes affect thermal conduction. In particular, the thermal conductivity of post-perovskite has been estimated to be 20\%--60\% higher than that of bridgmanite \citep{ohta2012lattice, okuda2020thermal, wang2023thermal}. Post-perovskite is expected to predominantly be stable in the colder parts of the lowermost mantle due to the large and positive Clapeyron slope of the bridgmanite to post-perovskite transition. The higher conductivity therefore has the potential to not only enhance the total CMB heat flux, but also its spatial variations.  

Another factor affecting the use of our results in geodynamo simulations is the balance of the thermal and compositional driving forces for outer core convection. The present-day geodynamo is in part driven by the release of light elements at the inner-core boundary \citep{nimmo2015energetics, landeau2022sustaining}. These light elements are not effectively accommodated by the mantle, making a vanishing flux across the CMB the most realistic boundary condition for chemical convection \citep{wicht2019advances}.
In other words, the heterogeneous CMB heat flux would only affect thermal and not chemical buoyancy, reducing the impact on outer core convection (and therefore the effective value of $q*$) the more of the driving force is contributed by chemical convection.

Finally, it is important to discuss the reference frame of our models, specifically when discussing the equatorial heat flux or the distribution of high and low heat flux patches. 
We here provide the CMB heat flux pattern in the reference frame of the plate reconstruction \citep[a palaeomagnetic reference frame derived from][]{tetley2018constraining}. 
On geological timescales, the geodynamo coincides with the Earth's spin axis, 
\cite[i.e.][]{van2015paleolatitude}. Not taking into account small deviations on the order of several degrees between the magnetic north pole and the spin axis derived from non-dipole field components, our models using this reference frame therefore provide CMB heat flux patterns with respect to the Earth's spin axis. Note, however, that paleolongitude is generally not well-constrained in plate reconstructions.

\section{Conclusions}

In this study, we quantify the spatial variability and absolute heat flux changes at the CMB on long (billion-year) time scales using state-of-the-art numerical methods that improve the accuracy of heat flux computations in geodynamic models.
We find that there are characteristic changes in the heat flux pattern throughout the supercontinent cycle, and that subduction history drives the changes in these patterns. As long as there is a layer of intrinsically dense material at the base of the mantle, cold material accumulates below the location of subduction zones at the Earth's surface---causing a large heat flux out of the core---and pushes hot material to the areas between these cold zones, which are characterised by a CMB heat flux close to zero. This behavior occurs for all combinations of material properties we have tested, including an increased viscosity of the intrinsically dense material. 
The number and morphology of these hot and intrinsically dense thermochemical structures depends on subduction history and the stage of the supercontinent cycle. Stable subduction zones such as in Earth's immediate past lead to coherent and stable piles. Conversely, at times in Earth's history when subduction location changes frequently, thermochemical piles can fork or split up.
If subduction zones are located predominantly in the supercontinent hemisphere during supercontinent assembly, a degree-1 pattern can develop, with high CMB heat flux (cold material) located beneath the supercontinent and low heat flux (hot thermochemical structures) located beneath the superocean. 
In return, a ring of subduction zones around the supercontinent during its break-up tends to lead to a degree-2 pattern, with two hot thermochemical structures separated by this girdle of cold material.
Only models without a dense basal layer feature a significantly different heat flux pattern with areas of high and low heat flux alternating on much shorter spatial scales. This is because without a dense layer, cold remnants of subducted slabs can spread out along the core-mantle boundary more easily.

The amplitude of spatial heat flux variations at a given point in time is primarily affected by the material properties of the mantle. A thermal conductivity that increases with depth, a viscosity reduction in the lowermost mantle as expected for the post-perovskite phase, and the presence of an intrinsically dense layer all increase the amplitude of spatial heat flux variations. Together, these factors can increase this amplitude by a factor of 3. For a given set of material properties, the amplitude of heat flux variations (as characterized by $q*$) only varies moderately (by 30-50\%) over time. 
Our minimum estimate is $q*\geq2$ (and likely $q*\geq6$), but $q*$ might be $>30$ depending on the adiabatic heat flux out of the core. Lowermost mantle material properties, in particular the viscosity, strongly affect the timing of maxima and minima in the total heat flux, equatorial heat flux and the amplitude of spatial heat flux heterogeneity. 
Therefore, the current uncertainties both in plate motion history and lowermost mantle properties prohibit better constraints on the temporal evolution of the CMB heat flux and connecting them to paleomagnetic observations.
However, we hope that our results will serve as a tool in future studies to better quantify the effect of CMB heat flux heterogeneity on the geodynamo and improve our understanding of the connection between past mantle flow and changes in the magnetic field behavior.

\begin{acknowledgments}
The authors would like to acknowledge Joseph G. Meert, Andrew J. Biggin, Chris J. Davies, and Richard K. Bono for helpful discussions and suggestions that improved our manuscript.
The authors acknowledge support by NSF award EAR 2054605---Understanding the influence of mantle dynamics on the generation of Earth's magnetic field throughout the plate tectonics cycle. JD and RG were also partially supported by NSF awards EAR 1925677 and 2149126. CJS was also partially supported by NSF awards EAR 2016763 and 2237807.
ASPECT is hosted by the Computational Infrastructure for Geodynamics (CIG) which is supported by the National Science Foundation awards EAR 1550901 and 2149126.
The authors also acknowledge the University of Florida Research Computing (https://www.rc.ufl.edu/) and the Extreme Science and Engineering Discovery Environment (XSEDE) (award MCA08X011), which is supported by National Science Foundation award ACI 1548562, for providing computational
resources and support that have contributed to the research results reported in this publication.
The maps in the study were made using SubMachine~\citep{hosseini2018submachine}, Cartopy \citep{Cartopy} and
GPlates~\citep{muller2018gplates}, and the 3-D images, cross sections and videos were created using ParaView \citep{ahrens200536}.

Author contributions: 
Conceptualization: JD, RG, CS, DT (equal);
Methodology: JD, RG (lead), DT (supporting);
Investigation: JD (lead), FL (supporting);
Formal analysis: JD (lead), all (supporting);
Validation: RG;
Visualization and Writing -- original draft: JD (lead), RG \& DT (supporting);
Writing -- review \& editing: all.

\end{acknowledgments}

\begin{dataavailability}

The input data and plate reconstruction files to reproduce the models in this study, and the output data necessary to reproduce the figures are available from Zenodo \citep{dannberg2023changes} under the doi \href{https://zenodo.org/record/8408548}{10.5281/zenodo.8408548} under a creative commons license. We used the modeling software ASPECT \citep{bangerth2022aspect} to compute our models, which is available via a GPLv2 or newer license and developed openly at \url{www.github.com/geodynamics/aspect}.
The specific version of ASPECT being used in this study is archived on Zenodo \citep{dannberg2023changes} as well. 
\end{dataavailability}

\bibliographystyle{gji}
\bibliography{main}

\appendix
\counterwithin{figure}{section}
\counterwithin{table}{section}
\renewcommand\thefigure{\thesection\arabic{figure}}
\renewcommand\thetable{\thesection\arabic{table}}

\section{Supporting videos}

Each video shows the evolution of one of the models listed in Table~\ref{tab:models}. White-to-blue colors represent subducted slabs (material that is at least 200~K colder than the mantle adiabat) below 670 km depth, with deeper hues of blue indicating greater pressure/depth. Yellow-to-orange colors highlight the distribution of intrinsically dense basaltic material at the base of the mantle, with lighter yellow indicating greater pressure/depth. The red stripes at the Earth's surface show regions of high strain rate, illustrating the location of plate boundaries. The left panel displays the African hemisphere, the right panel the Pacific hemisphere. 

\textbf{Video~S1}: \href{https://youtu.be/0ZxLYyc_7tM}{Thermochemical model}. 

\textbf{Video~S2}: \href{https://youtu.be/QT18stc9iyw}{Thermal model}.

\textbf{Video~S3}: \href{https://youtu.be/G7gK6ME04dA}{p-T-dependent conductivity model}.

\textbf{Video~S4}: \href{https://youtu.be/8kRzHXiD6tQ}{Weak post-perovskite model}.

\textbf{Video~S5}: \href{https://youtu.be/eYPbYeZKjvs}{Strong basalt model}.

\FloatBarrier
\section{Additional Figures}

\begin{figure}
    \begin{center}
        \includegraphics[width=\columnwidth]{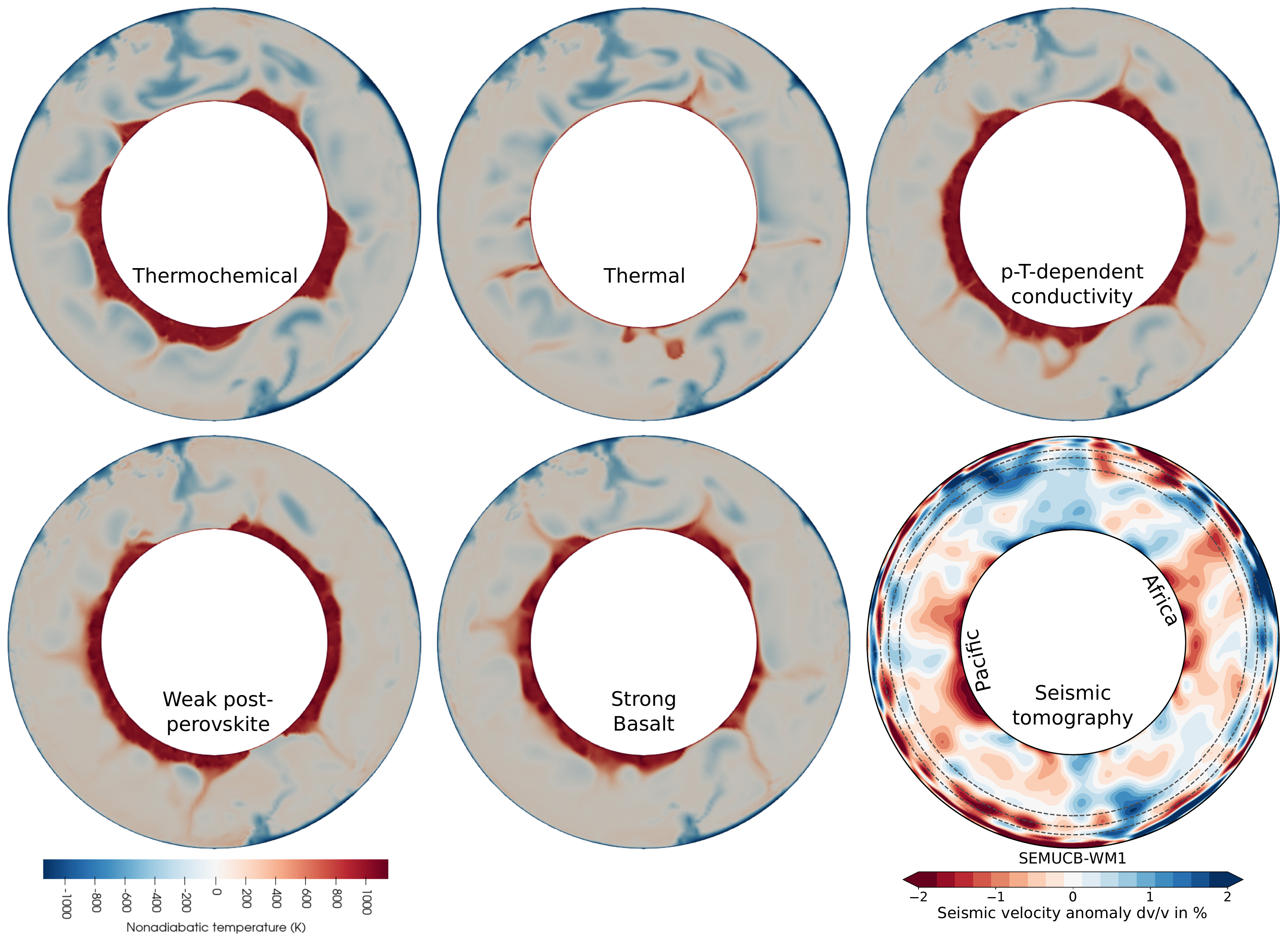}
        \caption{Temperature distribution (shown as the difference compared to the initial mantle adiabat) in an equatorial slice through all models at the end of the simulation (present-day). This is the same as Figure~\ref{fig:temperature_eq_slice}, but with a colorscale that is symmetric about a temperature anomaly of zero.}
        \label{fig:temperature_eq_slice2}
    \end{center}
\end{figure}
\FloatBarrier

\section{The consistent boundary flux method for heat flux computation}
\label{appx:CBF}

We here summarize a simplified numerical approach of the CBF method: To avoid the inaccuracies associated with gradient-based approached outlined in Section~\ref{sec:cbf}, the CBF method reformulates the underlying equation (in this case the energy equation~\eqref{eq:aspect-system-3}) to solve for the necessary flux across the boundary of the domain that satisfies the numerically computed solution of the original problem. In other words, we ask the question: If the temperature was not prescribed at the boundary, what heat flux would be consistent with the temperature solution that was just computed? This problem requires solving an additional equation, but since this equation only needs to be solved on the boundary of the domain it is computationally cheap compared to the original equation.

In mathematical and general terms we want to solve~\eqref{eq:aspect-system-3} in the domain $\Omega$ given the initial condition $T(\boldsymbol{x},0) = g(\boldsymbol{x})$ and boundary conditions:
\begin{align}
    T = T_0 \quad \text{on} \quad \delta \Omega_1, \\
    \frac{k}{\rho C_p} \mathbf{n} \cdot \nabla T = -q_2 \quad \text{on} \quad \delta \Omega_2.
\end{align}
Here, $\partial \Omega_1$ is the part of the boundary where Dirichlet boundary conditions are prescribed (prescribed temperature), $\partial \Omega_2$ is the part of the boundary where Neumann boundary conditions are prescribed (prescribed heat flux), $q_2$ is any prescribed boundary heat flux on $\partial \Omega_2$, and $\mathbf{n}$ is the unit vector normal to the boundary. 
$T$ is temperature, $k$ thermal conductivity, $\rho$ density, $C_p$ specific heat capacity, and $g(\boldsymbol{x})$ is the initial temperature in dependence of the location $\boldsymbol{x}$. Note that~\citet{gresho1987consistent} and ASPECT define the normal of the boundary $\mathbf{n}$ in outwards direction, so that in our case heat flux into the mantle and out of the core results in negative values for the heat flux. However, we have flipped the sign of the output in all figures and in the main text for ease of understanding.

Reformulating equation (30) of~\citet{gresho1987consistent} into the weak form of the consistent boundary flux method for our problem yields:
\begin{equation}
\label{eq:CBF-1}
    \int_\Omega \Gamma_i \left[ \frac{\partial T}{\partial t} + \mathbf{u} \cdot \nabla T \right] + \int_\Omega \frac{k}{\rho C_p} \nabla \Gamma_i \cdot \nabla T = \int_\Omega \Gamma_i S + \int_{\partial \Omega_2} q_2 - \int_{\partial \Omega_1} \Gamma_i q_1
\end{equation}
In this equation we use the notation of~\eqref{eq:aspect-system-3}, $\Gamma_i$ are those discretized temperature basis functions that are non-zero on the boundary, $\mathbf u$ velocity, $S$ all heat sources (all right-hand side terms of~\eqref{eq:aspect-system-3}), and $q_1$ the heat flux through boundaries  with Dirichlet conditions (which we want to solve for). Compared to~\citet{gresho1987consistent} we additionally simplify the equation to assume we solve the Stokes equation without spurious numerical velocity divergence, and only considering Dirichlet or Neumann boundary conditions. In order to solve this equation we utilize the same type of finite elements that are used for the temperature equation ($Q_2$) on the boundary and expand 
\begin{align}
q_1 =& \sum_{j=1}^{M} q_{1j}\Gamma_j \mid _{\partial \Omega_1} \qquad \text{which, when inserted into~\eqref{eq:CBF-1} yields} \\
\sum_{j=1}^{M} q_{1j} \int_{\partial \Omega_1} \Gamma_i \Gamma_j =& 
        -\int_\Omega \Gamma_i \left[ \frac{\partial T}{\partial t} - \mathbf{u} \cdot \nabla T \right] 
        - \int_\Omega \frac{k}{\rho C_p} \nabla \Gamma_i \cdot \nabla T \notag \\
        & + \int_\Omega \Gamma_i S + \int_{\partial \Omega_2} q_2 \quad i=1,2,...,M.
\end{align}
Making use of a Gauss--Lobatto--Legendre quadrature for the quadrature points on the boundary (which are colocated with the finite element support points of our chosen element) ensures that the resulting linear system has only entries on the main diagonal and is therefore easily inverted.

\begin{figure}
\noindent\includegraphics[width=\textwidth]{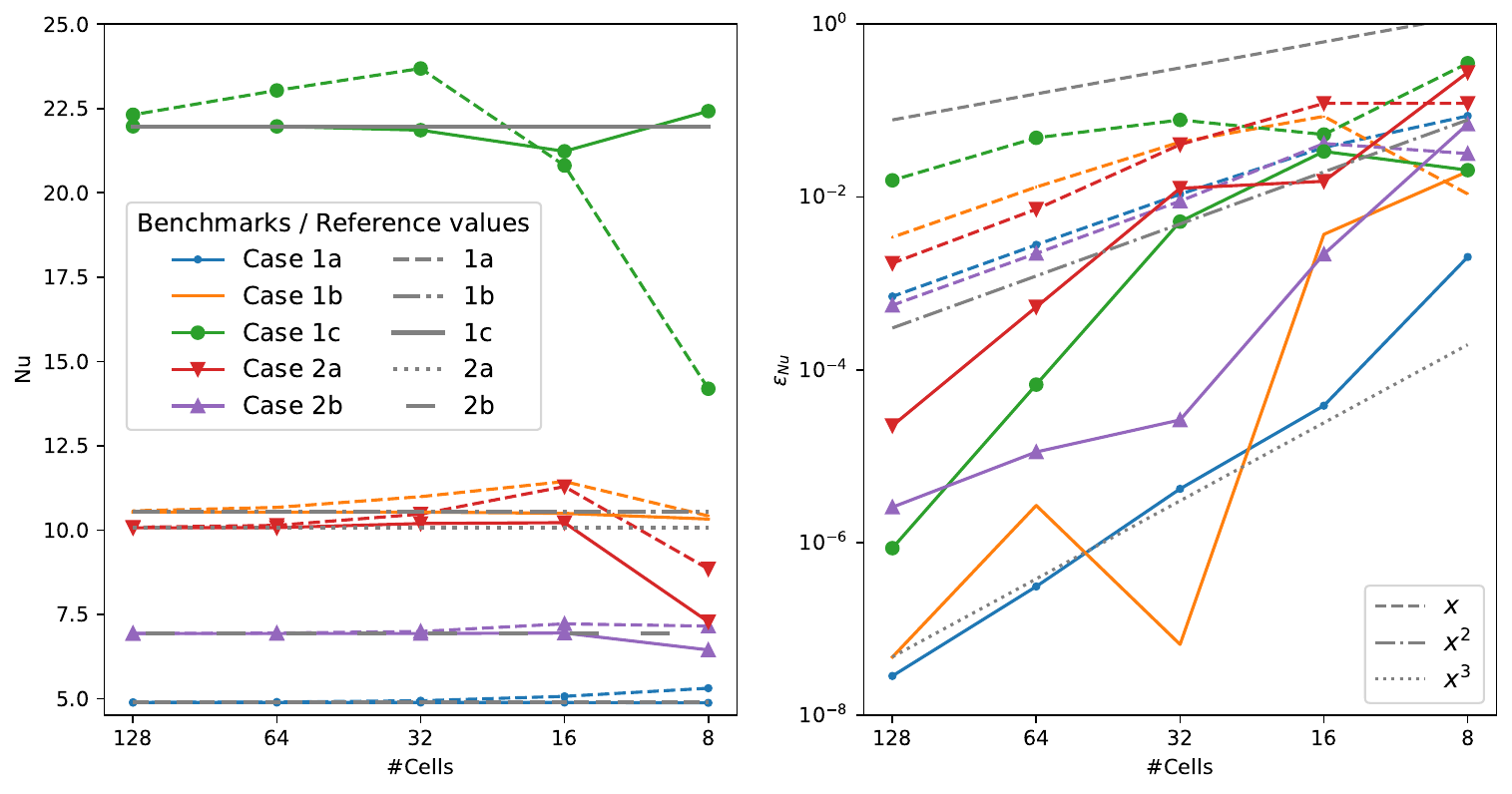}
\noindent\includegraphics[width=\textwidth]{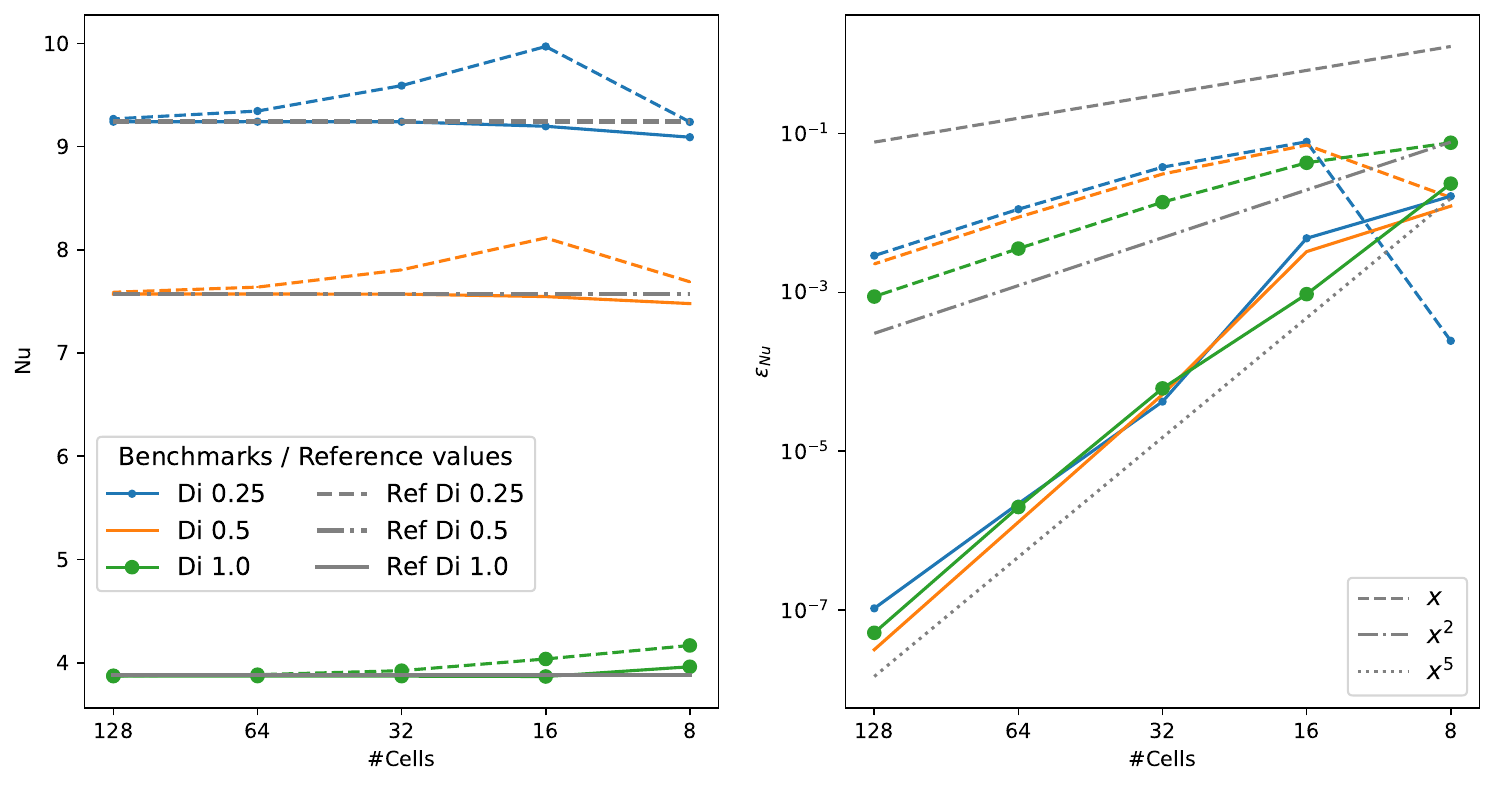}
\caption{Nusselt number (boundary heat flux) over the number of cells (resolution) for the benchmark models of~\citet{blankenbach1989benchmark} (top panels) and~\citet{king2010community} (bottom panels). We plot results of the CBF heat flux method (solid lines) and a gradient based heat flux method (dashed lines) compared to the reference results reported in the publications (gray lines). Note that the bottom row only shows the case of $Ra=10^5$ and the anelastic liquid approximation for different dissipation numbers to improve the visibility of the figure and that we used the case ``RefVT'' as reference value. The right column shows the respective convergence rate of the relative error $\epsilon_{Nu}$ over number of cells. The numeric values of the results are given in Tables~\ref{tab:cbf_benchmark} and \ref{tab:cbf_king_benchmark}.}
\label{fig:full_cbf_benchmark}
\end{figure}

We measure the relative error $\epsilon$ of our CBF implementation as the difference between the computed Nusselt number \textit{Nu} and a reference value \textit{Nu}$_\text{ref}$ computed using a Richardson extrapolation of the model results of increasing resolution. The Richardson extrapolation value is assumed to be close to the exact value, which is supported by the observation that it is always very close (with a relative difference of $10^{-3}-10^{-4}$ with only a few exceptions around $10^{-2}$) to the values reported in~\citet{blankenbach1989benchmark} and~~\citet{king2010community}, see Table~\ref{tab:cbf_benchmark} and Table~\ref{tab:cbf_king_benchmark}. Comparing the relative error of CBF ($\epsilon_\text{CBF}$) to the relative error of a heat flux computed as the thermal conductivity times the temperature gradient at the model boundary ($\epsilon_\text{grad}$) we notice that CBF results in a significantly smaller error and a better convergence rate with increasing resolution. In particular, at very coarse resolutions of 8 or 16 cells the CBF error is significantly smaller than the gradient error for well resolved simulations (Blankenbach case 1a), while for underresolved simulations both errors are of comparable magnitude. For fine resolutions (64 cells) the CBF error is uniformly smaller by 1--4 orders of magnitude. We observe that the convergence rate of the CBF method (right panels of Figure~\ref{fig:full_cbf_benchmark}) is better by at least one order (in low $Ra$ number cases or when the flow is sufficiently resolved) and up to three orders (in high $Ra$ number cases). In our opinion this variability in the convergence rate of the CBF method is caused by the larger number of input properties that are used to compute the heat flux. While the gradient based method only depends on the derivative of the temperature (and therefore converges with one order less than the temperature solution) the CBF heat flux depends on temperature, velocity, and energy source and sink terms and depending on which of these terms dominate, the remaining error will converge with the convergence rate of that property.

\begin{table}
\caption{Nusselt number and root mean square velocity of benchmark cases in~\citet{blankenbach1989benchmark} (B89) using the consistent boundary flux method ($\textit{Nu}_\text{CBF}$) and a temperature gradient based computation ($\textit{Nu}_\text{grad}$). Note that both methods use the same solution to compute the postprocessed heat flux. The ASPECT reference result (AS) was computed using a Richardson extrapolation and the relative difference between each resolution and the reference value $\epsilon$ is computed as $\epsilon = \lvert(\textit{Nu} - \textit{Nu}_\text{AS}) / \textit{Nu}_\text{AS}\rvert$. The values are plotted in Figure~\ref{fig:full_cbf_benchmark} (top row).}
\label{tab:cbf_benchmark}
\centering
\setlength\extrarowheight{-3pt}
\begin{tabular}{l c c c c c c}
\hline
Case & Cells & $\textit{Nu}_\text{CBF}$ & $\epsilon_{\textit{Nu}\text{CBF}}$ &          $\textit{Nu}_\text{grad}$ & $\epsilon_{\textit{Nu}\text{grad}}$ &         V$_\text{rms}$  \\
\hline
1a & 8 &  4.8745247 & 0.0020237002 &  5.3088532 & 0.086897697 &  42.849993 \\
1a & 16 &  4.8845974 & 3.8510286e-05 &  5.0673529 & 0.037454603 &  42.865678 \\
1a & 32 &  4.8844298 & 4.1970275e-06 &   4.937124 & 0.010792431 &  42.865035 \\
1a & 64 &  4.8844108 & 3.1119423e-07 &  4.8980783 & 0.0027984858 &  42.864962 \\
1a & 128 &  4.8844094 & 2.8662627e-08 &  4.8878581 & 0.00070609152 &  42.864957 \\
1a & AS &  4.8844093 &          0 &  4.8842346 &          0 &  42.864956 \\
1a & B89 &   4.884409 &          0 &   4.884409 &          0 &  42.864947 \\
1b & 8 &  10.325881 & 0.019748694 &  10.419534 & 0.010858103 &  195.27937 \\
1b & 16 &  10.494936 & 0.0037000404 &  11.438417 & 0.08586602 &  193.08784 \\
1b & 32 &  10.533913 & 6.6452045e-08 &  10.990004 & 0.043297504 &  193.21456 \\
1b & 64 &   10.53394 & 2.6960544e-06 &  10.671379 & 0.013049938 &  193.21479 \\
1b & 128 &  10.533912 & 4.7465747e-08 &  10.569992 & 0.0034251378 &  193.21455 \\
1b & AS &  10.533912 &          0 &  10.522674 &          0 &  193.22175 \\
1b & B89 &  10.534095 &          0 &  10.534095 &          0 &  193.21454 \\
1c & 8 &  22.422669 & 0.020486746 &  14.190277 & 0.35418081 &  889.26535 \\
1c & 16 &  21.231585 & 0.033721139 &  20.810348 & 0.052892183 &  839.48773 \\
1c & 32 &  21.858433 & 0.0051924078 &  23.687901 & 0.078069237 &  833.32105 \\
1c & 64 &   21.97104 & 6.7493387e-05 &   23.03726 & 0.048457653 &  833.97214 \\
1c & 128 &  21.972504 & 8.6471635e-07 &  22.315716 & 0.015619171 &  833.99031 \\
1c & AS &  21.972523 &          0 &    29.6585 &          0 &  833.99083 \\
1c & B89 &  21.972465 &          0 &  21.972465 &          0 &  833.98977 \\
2a & 8 &  7.2792059 & 0.27683732 &  8.8459437 &  0.1211876 &     439.26 \\
2a & 16 &  10.218989 & 0.015219437 &  11.288971 & 0.12151825 &  464.97188 \\
2a & 32 &  10.193035 & 0.012641061 &  10.472918 & 0.040446431 &  475.81913 \\
2a & 64 &  10.071167 & 0.0005339172 &  10.138604 & 0.0072335384 &  480.16472 \\
2a & 128 &   10.06602 & 2.258143e-05 &  10.083072 & 0.0017165563 &  480.41987 \\
2a & AS &  10.065793 &          0 &  10.072009 &          0 &  480.43579 \\
2a & B89 &     10.066 &          0 &     10.066 &          0 &   480.4334 \\
2b & 8 &  6.4451282 & 0.069930429 &  7.1503442 & 0.031836354 &  168.57551 \\
2b & 16 &  6.9448381 & 0.0021806227 &  7.2182807 & 0.041639984 &  169.97689 \\
2b & 32 &  6.9299095 & 2.6337257e-05 &  6.9918953 & 0.008971251 &  171.55515 \\
2b & 64 &  6.9296493 & 1.1216892e-05 &  6.9451426 & 0.0022245667 &  171.74454 \\
2b & 128 &  6.9297091 & 2.575859e-06 &  6.9336044 & 0.00055953719 &  171.75313 \\
2b & AS &   6.929727 &          0 &  6.9298239 &          0 &  171.75354 \\
2b & B89 &     6.9299 &          0 &     6.9299 &          0 &    171.755 \\
\hline
\end{tabular}
\end{table}

\begin{table}
\caption{Nusselt number and other benchmarks properties of~\citet{king2010community}  (models CU, UM, VT) using the consistent boundary flux method ($\textit{Nu}_\text{CBF}$) and a temperature gradient based computation ($\textit{Nu}_\text{grad}$). Note that both methods use the same solution to compute the postprocessed heat flux. The ASPECT reference result (AS) was computed using a Richardson extrapolation and the relative difference between each resolution and the reference value $\epsilon$ is computed as $\epsilon = \lvert(\textit{Nu} - \textit{Nu}_\text{AS}) / \textit{Nu}_\text{AS}\rvert$. The values are plotted in Figure~\ref{fig:full_cbf_benchmark} (bottom row).}
\label{tab:cbf_king_benchmark}
\centering
\resizebox{\textwidth}{!}{%
\begin{tabular}{rlrrrrrrrr}
\hline
Di & Cells & $\textit{Nu}_\text{CBF}$ & $\epsilon_{\textit{Nu}\text{CBF}}$ & $\textit{Nu}_\text{grad}$ & $\epsilon_{\textit{Nu}\text{grad}}$ & V$_\text{rms}$ & $\langle T \rangle$ & Phi & W \\
\hline
      0.25 & 8 &     9.0912 & 0.016267465 &   9.239266 & 0.0002456756 &  181.50236 &  0.5352244 &  2.1292708 &  2.1301381 \\
      0.25 & 16 &  9.1971506 & 0.0048028562 &  9.9699623 & 0.078820863 &  177.87587 & 0.53189775 &  2.0495161 &  2.0488624 \\
      0.25 & 32 &  9.2411469 & 4.2151e-05 &  9.5908511 & 0.037798335 &  178.07174 & 0.53213716 &  2.0524867 &   2.051655 \\
      0.25 & 64 &  9.2415566 & 2.1879479e-06 &  9.3446185 & 0.01115422 &  178.07524 & 0.53215621 &  2.0525181 &  2.0516818 \\
      0.25 & 128 &  9.2415374 & 1.049609e-07 &  9.2683964 & 0.0029064485 &  178.07506 & 0.53215643 &  2.0525123 &   2.051676 \\
      0.25 & AS &  9.2415364 &          0 &  9.2342232 &          0 &  178.07505 & 0.53215644 &  2.0525111 &  2.0516744 \\
      0.25 & CU &       9.21 &          0 &       9.21 &          0 &      178.2 &     0.5319 &     2.0503 &      2.054 \\
      0.25 & UM &      9.196 &          0 &      9.196 &          0 &    178.229 &      0.532 &      2.041 &      2.051 \\
      0.25 & VT &     9.2428 &          0 &     9.2428 &          0 &   179.7523 &     0.5318 &     2.0518 &     2.0519 \\
       0.5 & 8 &  7.4793276 & 0.012208946 &  7.6903974 & 0.015666949 &  157.55928 & 0.55294319 &  3.3452266 &  3.3460926 \\
       0.5 & 16 &   7.547057 & 0.0032639524 &  8.1151862 & 0.071768591 &   154.9847 & 0.54764289 &  3.2336902 &  3.2323114 \\
       0.5 & 32 &  7.5713791 & 5.1739547e-05 &  7.8061962 & 0.030960427 &  155.11626 & 0.54800162 &  3.2367771 &   3.235111 \\
       0.5 & 64 &  7.5717806 & 1.2810742e-06 &  7.6388628 & 0.0088607937 &  155.11968 & 0.54803796 &  3.2368259 &   3.235151 \\
       0.5 & 128 &  7.5717711 & 3.169668e-08 &  7.5890635 & 0.0022838316 &  155.11958 & 0.54803856 &  3.2368182 &  3.2351431 \\
       0.5 & AS &  7.5717709 &          0 &  7.5679636 &          0 &  155.11958 & 0.54803857 &  3.2368168 &  3.2351411 \\
       0.5 & CU &       7.55 &          0 &       7.55 &          0 &      155.1 &     0.5472 &      3.233 &     3.2392 \\
       0.5 & UM &      7.532 &          0 &      7.532 &          0 &    155.304 &      0.548 &      3.221 &      3.233 \\
       0.5 & VT &     7.5719 &          0 &     7.5719 &          0 &   156.5589 &     0.5472 &     3.2344 &     3.2346 \\
         1 & 8 &  3.9610152 & 0.023444568 &  4.1672459 & 0.076730332 &  98.001745 & 0.53895275 &   3.026465 &  2.8341141 \\
         1 & 16 &  3.8665981 & 0.00095086446 &  4.0363703 & 0.04291476 &  84.318518 & 0.52934166 &  2.7563994 &  2.7713728 \\
         1 & 32 &  3.8700388 & 6.186377e-05 &  3.9232271 & 0.013680903 &  84.363966 & 0.52975972 &  2.7519461 &  2.7691969 \\
         1 & 64 &  3.8702705 & 1.9817697e-06 &  3.8840794 & 0.0035659555 &  84.367559 & 0.52980247 &   2.751865 &  2.7691556 \\
         1 & 128 &   3.870278 & 5.1675872e-08 &  3.8737095 & 0.00088657709 &  84.367814 & 0.52980371 &  2.7518642 &  2.7691555 \\
         1 & AS &  3.8702782 &          0 &  3.8699727 &          0 &  84.367833 & 0.52980374 &  2.7518642 &  2.7691555 \\
         1 & CU &       3.88 &          0 &       3.88 &          0 &       84.6 &     0.5294 &     2.7652 &     2.7742 \\
         1 & UM &      3.857 &          0 &      3.857 &          0 &     84.587 &       0.53 &      2.742 &      2.765 \\
         1 & VT &      3.878 &          0 &      3.878 &          0 &    85.5803 &     0.5294 &      2.761 &     2.7614 \\
\hline
\end{tabular}}

\end{table}

\label{lastpage}
\end{document}